\documentclass [peerreview, 12pt]{IEEEtran}

\usepackage{authblk}
\usepackage{algorithm}
\usepackage{algorithmic}
\usepackage{graphicx,amsmath,amssymb,latexsym,epsfig}
\usepackage{setspace}
\newtheorem{theorem}{Theorem}
\newtheorem{problem}{Problem}
\newtheorem{corollary}{Corollary}

\newtheorem{lemma}{Lemma}
\newtheorem{prop}{Proposition}

\newcommand{\enp} {\hfill \rule{2.2mm}{2.6mm}}

\begin{document}
\title{Optimal Packet Scheduling on an Energy Harvesting Broadcast Link} 
\author{Mehmet Akif Antepli}
\author{Elif Uysal-Biyikoglu}
\author{Hakan Erkal}
\affil{Dept. of Electrical and Electronics Eng., METU, Ankara 06531 Turkey\\
{akifantepli@gmail.com, elif@eee.metu.edu.tr, herkal@ieee.org}}
\bibliographystyle{IEEE}
\maketitle

\def\eg{{e.g.}}
\def\ie{{i.e.}}

\doublespacing
%\singlespacing

\begin{abstract}
The minimization of transmission completion time for a given number of bits per user in an energy harvesting communication system, where energy harvesting instants are known in an offline manner is considered. An achievable rate region with structural properties satisfied by the 2-user AWGN Broadcast Channel capacity region is assumed. It is shown that even though all data are available at the beginning, a non-negative amount of energy from each energy harvest is deferred for later use such that the transmit power starts at its lowest value and rises as time progresses. The optimal scheduler ends the transmission to both users at the same time. Exploiting the special structure in the problem, the iterative offline algorithm, FlowRight, from earlier literature, is adapted and proved to solve this problem. The solution has polynomial complexity in the number of harvests used, and is observed to converge quickly on numerical examples.
\end{abstract}

\begin{IEEEkeywords}
Packet scheduling, energy harvesting, AWGN broadcast channel, flowright, energy-efficient scheduling.
\end{IEEEkeywords}

\section{Introduction}
Since its formulation a decade ago~\cite{PUE01}, the problem of energy-efficient packet transmission scheduling has drawn considerable interest from the research community~\cite{BeGa02,NuSr02,ZaMo09}. The basic offline problem is to assign transmission durations (equivalently, code rates) to a set of packets whose arrival times are known beforehand, so that they are all transmitted within a given time window with minimum total energy. Recently, the problem has been recast with a formulation where the goal is to minimize the \emph{time} by which all packets are transmitted, given that energy is harvested at certain known instants~\cite{YaU2010}.

In this paper, we extend the formulation in~\cite{YaU2010} to a multiuser scenario with one sender and multiple receivers. In particular, we assume an AWGN Broadcast Channel where the sender gets replenished with arbitrary amounts of energy at arbitrary points in time. The harvested energy becomes instantly available for use, and the transmission power can be changed at any time by the sender. The choices of power level and the rates to individual receivers across time is called a \emph{schedule}. The sender needs to transmit a certain number of bits to each receiver. We consider the case that these bits are available at the beginning of transmission. The goal is to find a schedule that achieves the unique minimum time $T^{\rm{opt}}$, by which the data of all users can been transmitted using the given sequence of harvests. Throughout the paper, we focus on the \emph{offline} problem, where the energy harvesting times as well as packet arrival times are known in advance. The online version of the problem in which the times of energy harvests are not known a priori and decisions need to be made in real-time as the harvests occur, is interesting yet analytically less tractable and left outside the scope of this paper.

It is well known that both with optimal and practical coding schemes, the energy per bit increases with the transmission rate, in other words, transmitting fast is inefficient in terms of energy~\cite{UPE02}. This is the root of the sender's dilemma: it will pay off for the sender to slow down, yet it needs to minimize the overall transmission duration. Interestingly, it turns out that even if all packets were available in the beginning, the optimal schedule starts slowly, deferring some of the harvested energy for future use. More precisely, we will show that in the optimal schedule the transmission power is non-decreasing in time, similarly to the point-to-point schedule~\cite{YaU2010}.

In the point-to-point problem, determining power levels determines the schedule, as transmission rate is a function of average power. In the broadcast problem, however, there is no one-to-one correspondence between the transmission power and the rate point. For example, with optimal coding, there is a continuum of rates on the boundary of the capacity region corresponding to a certain average power constraint. Hence, the rates and the power have to be determined together. We observe that in the optimal schedule, the average rates used by the users are proportional to their numbers of bits, i.e. the schedule always continuously transmits to all users at the same time and finishes transmission to all users at the same time. Having made this observation, we can exploit the mathematical similarities between this problem and the problem in~\cite{Eu04}, and show that the solution is found by the algorithm FlowRight, defined in~\cite{Eu04} and adapted here to work with different parameters. 

In the next section, we make observations about the two-user AWGN broadcast channel capacity region. The statement of the problem as a cost minimization problem, as well as its solution will use certain structural properties of the AWGN capacity region, such as the monotonicity and convexity of the average power with respect to the rate pair. Of course, this specific rate region can only be approached under optimal coding as blocklengths and the number of information bits go to infinity. For example, in the single user AWGN channel the numerical value of the minimum energy per bit corresponding to a given reliability monotonically decreases with the number of information bits~\cite{PPVISIT2010}. However, the basic structural properties of the rate region will be satisfied by the achievable rate regions of many suboptimal practical coding schemes as well as finite blocklength optimal coding schemes. 

We define the problem in Section \ref{sec:probdef}. In Section \ref{sec:solution} we explore the properties of the optimal solution. This is followed by the description of the modified FlowRight algorithm, and the proof of its convergence and optimality of the resulting schedule. The complexity of the iterative algorithm is  analyzed in Section \ref{sec:AlgorithmComplexity}. The implementation of this algorithm is discussed, followed by a numerical example in Section \ref{sec:Example}. Section~\ref{conclusion} summarizes our conclusions and outlines further directions.

\section{Broadcast Channel}
\label{sec:bc}

\iffalse
\begin{figure}[hbt]
\centering \includegraphics[scale=0.3]{Broadcast_Channel.pdf}
\caption{A two-user broadcast link.}
\label{fig:BroadcastChannel}
\end{figure}
\fi
Consider a discrete-time AWGN broadcast channel with one sender and two receivers. % as shown in Fig.~\ref{fig:BroadcastChannel}. 
The signal received by the $i^{th}$ user at time $k$ is given by
\begin{equation}
\label{eq:broadcastchannelmodel}
Y_i[k]=\sqrt{s_i}X[k]+Z_i[k],
\end{equation}
where $X[k]$ is the transmitted signal with average power constraint $P$, $\sqrt{s_i}$'s are the channel gains and the $Z_i[k]$'s are i.i.d. zero-mean Gaussian noise with variance $\sigma^2$.  The capacity region of the channel assuming $s_1$ and $s_2$ are constants and $s_1 > s_2 > 0$ , is the set of rate pairs $(r_1,r_2)$ such that~\cite{CoT91}
\begin{eqnarray}
r_1 &\leq & \frac{1}{2}\log_2\left(1+\frac{\alpha s_1 P}{\sigma^2}\right)\label{eq:Bcast_R1}\\
r_2 &\leq & \frac{1}{2}\log_2\left(1+\frac{(1-\alpha) s_2 P}{\alpha  s_2 P + \sigma^2}\right)\label{eq:Bcast_R2}
\end{eqnarray}
for some $0 \leq \alpha \leq 1$. Hence, the $1^{st}$ user is the \emph{stronger} user.

It is straightforward to show that, given $s_1$ and $s_2$, for any $P_1>P_2$, the capacity region corresponding to an average power constraint $P_1$ dominates the one corresponding to $P_2$. Therefore, given a rate pair $(r_1,r_2)$, there is a unique $P=g(r_1,r_2)$ (see~\cite{Eu04}) such that $(r_1,r_2)$ lies on the boundary of the rate region with power constraint equal to $P$. After replacing the inequalities in \eqref{eq:Bcast_R1} and \eqref{eq:Bcast_R2} by equalities, the function $g(r_1,r_2)$ is written as follows. \eqref{eq:Bcast_R1} can be written as as $\alpha s_1 P/\sigma^2 = 2^{2 r_1 }-1$. Hence, $\alpha =(\sigma^2/P s_1)(2^{2 r_1 }-1) $. After substituting into \eqref{eq:Bcast_R1} and rearranging the terms, we obtain
\begin{equation}
\label{eq:g}
g(r_1,r_2) =\sigma^2\left(\frac{(2^{2 r_2 }-1)}{s_2}+\frac{(2^{2 r_1 }-1)2^{2 r_2}}{s_1}\right).
\end{equation}

The function $g(r_1,r_2)$ is twice continuously differentiable and strictly convex in $r_1$ and $r_2$. Throughout the paper, it will be useful to express the $r_1$ and $r_2$ as a function of each other and the minimum power $P$. By algebraic manipulation of \eqref{eq:Bcast_R1} and \eqref{eq:Bcast_R2}, we obtain the following:
\iffalse
From \eqref{eq:Bcast_R1} and \eqref{eq:Bcast_R2}, the function $g(r_1,r_2)$  is strictly convex and continuously differentiable in $r_1$ and $r_2$.
\fi
\begin{eqnarray}
\label{eq:h1}
r_1= h_1(P,r_2)&=&\frac{1}{2} \log_{2}(\frac{s_1 (s_2 P + \sigma^2)}{s_2 \sigma^2
2^{2 r_2}}-\frac{s_1-s_2}{s_2})\\
r_2= h_2(P,r_1)&=&\frac{1}{2} \log_{2}(\frac{\frac{s_2 P}{\sigma ^2}+1}{\frac{s_2}{s_1} (2^{2 r_1}-1)+1}). 
\label{eq:h2}
\end{eqnarray}

The properties satisfied by these rate functions for the AWGN BC capacity region with $s_1>s_2$ summarized in the following will be used in the rest of the paper. 

\underline{Properties of the rate region:}
\begin{enumerate}
\item Nonnegativity: $h_1(P,r) \geq 0 , h_2(P,r)\geq 0$.
\item Monotonicity: $h_1(P,r)$, $h_2(P,r)$ are both monotone decreasing in $r$, and monotone increasing in $P$.
\item Concavity: $h_1(P,r)$ and $h_2(P,r)$ are concave in $P$ and $r$.
\item The rate of the user with the weaker channel satisfies the following: $\frac{\partial^2 h_2(P,r)}{\partial r \partial P}=0$, $\frac{\partial^2 h_2(P,r)}{\partial P \partial r }=0.$
\end{enumerate}

\begin{prop}
\label{prop:h}
The functions $h_1$ and $h_2$, defined in \eqref{eq:h1},\eqref{eq:h2} on $\Re^+\times \Re^+$ satisfy the Properties of the rate region given by (1)-(4).
\end{prop}

\noindent{\emph{Proof.}} See Appendix A.

\section{PROBLEM DEFINITION}
\label{sec:probdef}
\begin{figure}[htpb]
\centering \includegraphics[scale=0.6]{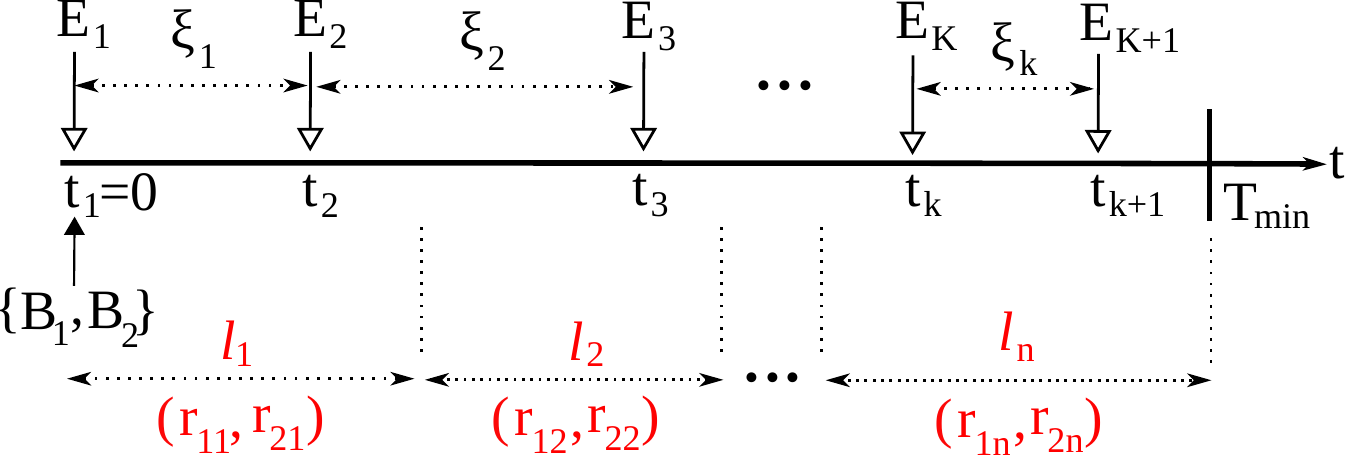}
\caption{System model with all of the bits of the two users are available at the beginning. Energies arrive at times $t_i$ where $i\in \{1,2,...,k+1$\}}
\label{fig:System_Model}
\end{figure}

Consider the broadcast link as described in the previous section, with a sender who needs to transmit $B_1<\infty$ and $B_2<\infty$ bits with a certain degree of reliability to users 1 and 2, respectively \footnote{Throughout the paper, two receivers will be considered for ease of exposition. However, the results can be generalized to more than two receivers.}. Also assume that at time $t_{1}$=0, sender has $E_1>0$ units of energy available and at times $t_2,...,t_{k+1}$, energies are harvested with amounts $E_2,..., E_{K+1}$, respectively, as depicted in Fig~\ref{fig:System_Model}. Inter-arrival times of energy harvests are named as \emph{epochs}, and marked with $\xi_i, i=1,...,k$. 

It will be assumed that the sender has the ability to change its rate pair at any time, according to the available energy and remaining number of bits. Such ideal adaptation, which has been used in previous literature (\eg,~\cite{YaU2010}, and references therein), may be approximated by using adaptive coding and modulation in a practical system. 

Starting at $t=0$, let $\{(r_{11}, r_{21}), (r_{12}, r_{22}),..., (r_{1n}, r_{2n}), \ldots\}$, be the successive pairs of rates used by the sender, and $\{l_1, l_2,..., l_n, \ldots\}$ be the respective durations for which these pairs are used. Here, $r_{ij}\geq 0$ is user $i$'s rate in the $j^{th}$ rate pair. By definition, at least one user's rate changes from one rate pair to the next one. We will refer to the sequence of rate pairs and durations as a {\emph{schedule}}. The problem of interest is to find an optimal {\emph{offline}} schedule, that is, a schedule that minimizes the overall transmission completion time of the $B_1$ and $B_2$ bits to their respective destinations, with complete knowledge of future energy harvesting instants and the amounts to be harvested. 

It will also be assumed that the problem is feasible; that is, sufficient energy will be harvested to transmit the given $B_1<\infty$ and $B_2<\infty$ bits in arbitrarily large but finite total time, $T$. Note that for any given $E$, there is a small enough rate (equivalently, long enough transmission duration) such that $B_1$ and $B_2$ bits can be transmitted with energy $E$, provided that the minimum energy per bit required for communication on the broadcast channel for the given finite amount of bits is satisfied~\cite{JKV09}. In the point-to-point case with infinite blocklengths, the well known limit for energy per bit is $-1.59$ dB. For sending finite amounts of data, the minimum energy per bit is higher even at nonvanishing values of error probability. However, the upper and lower bounds in~\cite{PPVISIT2010} on energy per bit come very close to the ideal limit at $B=10^3$ bits, and even at smaller numbers of bits. 

In order to define the two-user broadcast channel offline scheduling problem as an optimization problem, we will use the set of observations stated in Lemmas~\ref{lmm:CloserPowers}-\ref{lmm:Tmin_GeneralEpoch}. Lemma~\ref{lmm:Tmin_GeneralEpoch} will establish that in an optimal schedule the transmission to both users ends at the same time. Lemma~\ref{lmm:ConstantRate} will establish that in an optimal schedule the rates and power level do not change between energy harvests that are used. As a consequence of these two results, the problem reduces to Problem~\ref{pr:MultiuserScheduling}. 

We start by proving a more general result than Lemma~\ref{lmm:ConstantRate} which will be used in the proof of Lemma~\ref{lmm:ConstantRate} as well as Theorem 1 in Section~\ref{sec:probdef}. Specifically, we take a finite time window which is divided into two slots such that different power levels are used in each. We show that by using a more even distribution of power (reducing the difference of the power levels) as much as energy causality permits, at least the same amount of data can be transmitted in the same amount of time using the same amount of energy. In the special case when this time window is within (or all of) one epoch, all the energy that is used is available in the beginning hence the powers can be completely equalized. 

%%	=============================================
%%	=================	LEMMA	=================
%%	=============================================
\begin{lemma}
\label{lmm:CloserPowers}
Suppose that within a time window $(\tau_1, \tau_2)$, the sender changes its transmit power at point $\tau^*$ such that $\tau_1<\tau^*< \tau_2$. Keeping the total consumed energy in $(\tau_1, \tau_2)$ constant, the sender can send at least the same number of bits to the users within the same duration by bringing power levels closer to each other, if feasible (i.e., unless such a change requires energy to be used before its harvested.)
\end{lemma}

\begin{figure}[htpb]
\centering \includegraphics[scale=0.5]{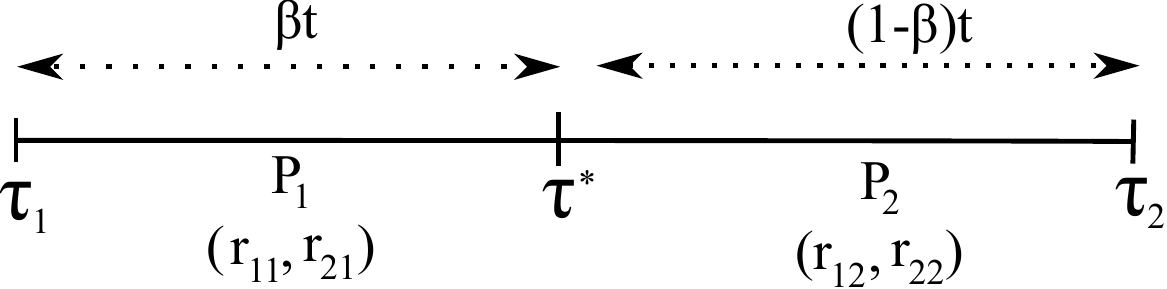}
\caption{Illustration of the transmission scheme used in Lemma~\ref{lmm:CloserPowers}.}
\label{fig:CloserPowers}
\end{figure}

{\emph{Proof.}} Let the total duration be $t=\tau_2-\tau_1$, and the lengths of the two slots $\beta t$ and $(1-\beta)t$, with power levels in the two slots $P_1$ and $P_2$, as illustrated in Fig.\ref{fig:CloserPowers}. Denote the rate pairs in the $1^{st}$ and $2^{nd}$ slots as ($r_{11}$, $r_{21}$) and ($r_{12}$, $r_{22}$), respectively. 

First, consider the case where the power level used in the first slot is smaller: $P_1 < P_2$. When $P_1\beta t$ is equal to the total energy available for use in $(t_1,\tau^{*})$, transferring energy from the second slot to the first is not feasible, and we stop. However, if it is possible to transfer some positive amount of energy $\Delta E$ from the second slot to the first, we shall show that we can only improve the allocation.

Let us denote the average rates for the \emph{stronger} and \emph{weaker} users as $\bar{r_{1}} \triangleq \beta r_{11} + (1-\beta) r_{12}$ and $\bar{r_{2}} \triangleq \beta r_{21} + (1-\beta) r_{22}$, respectively. We will show that keeping the total consumed energy and $\bar{r_{1}}$ constant, the sender can achieve an average rate $\bar{\bar{r_{2}}}$ for the second user such that $\bar{\bar{r_{2}}} \geq \bar{r_{2}}$ by changing $P_1$ to $P_1^{'}$ and $P_2$ to $P_2^{'}$ satisfying 
\begin{equation}
P_1 \leq P_1^{'} \leq P_2^{'} \leq P_2.
\label{eq:CloserPowers}
\end{equation}
While keeping the total energy constant, a certain amount of energy should be transferred from the $2^{nd}$ slot to the $1^{st}$ one in order to satisfy \eqref{eq:CloserPowers}. In this case, we have the following
\begin{equation}
P_1^{'}= P_1 + (1-\beta)\Delta P \mbox{~,~}P_2^{'}= P_2 - \beta \Delta P.
\label{eq:Del_P}
\end{equation}
Average rate belonging to the \emph{weaker} user over the whole duration $t$ is given by
\begin{eqnarray}
\bar{\bar{r_{2}}}&=& h_2(P_1^{'},\bar{r_{1}})\beta + h_2(P_2^{'},\bar{r_{1}})(1-\beta) \nonumber \\
&\geq & h_2(P_1,r_{11})\beta + h_2(P_2,r_{12})(1-\beta) \label{eq:Avg_2nd_User_Rate_1}= \bar{r_{2}} \nonumber
\end{eqnarray}
\eqref{eq:Avg_2nd_User_Rate_1} follows from the fact that
\begin{equation}
h_2(P_1^{'},\bar{r_{1}})\beta + h_2(P_2^{'},\bar{r_{1}})(1-\beta) - h_2(P_1,r_{11})\beta - h_2(P_2,r_{12})(1-\beta)\geq 0
\label{eq:f}
\end{equation}
for all $\beta=\{0,1\}$ with equality achieved at $\beta=0,1$. This is a consequence of the properties listed in Section~\ref{sec:bc}, and proved in Section~\ref{subsec:Appendixbeta} in the Appendix. 

In the remaining case, $P_1>P_2$, a similar argument holds where $P_2 \leq P_2^{'} \leq P_1^{'} \leq P_1$. In this case it is always possible to strictly improve the allocation by transferring a positive amount of energy from the first slot to the second, as energy can always be deferred for future use. 
\iffalse
We conclude that keeping the total consumed energy constant, one can find rate pairs such that at least the same number of bits can be transmitted to the users within the same duration by reallocating power levels closer to each other.
\fi
\enp

%We conclude that one can find a rate pair on the broadcast achievable region defined by the average power $P^{'}$ so that the sender can transmit at least the same number of bits to each user within the epoch as in the original, slotted allocation. Conversely, with this power allocation the transmission of the original number of bits to each user can only end sooner. Hence, it is not necessary to change either the rate pair or the power during an epoch. \enp \\

%%	=============================================
%%	================   COROLLARY  ===============
%%	=============================================
\begin{corollary}
\label{corollary:ConstantPower}
In a schedule that ends at $T^{\rm{opt}}$, power does not change {\emph{within}} epochs in $[0,T^{\rm{opt}}]$. 
\end{corollary}

{\emph{Proof.}} 
The claim is that the power does not change within epochs, of course with the exception of the last epoch. (In the last epoch that is used, the transmission ends and the power is reduced to zero at some point within the epoch.) By definition, no new energy or data is added during an epoch, so it is intuitive that the decision on power allocation does not change at a point during an epoch. To reach contradiction, suppose that the sender changes its power allocation during an epoch. From Lemma~\ref{lmm:CloserPowers}, the power levels can be allocated closer to each other so that at least the same number of bits can be transmitted to the users. Since this case is not limited by causality, this procedure can be continued until the power levels within the epoch are equalized, strictly improving the schedule, contradicting the optimality of the original schedule.   
\enp
%%	=============================================
%%	=================	LEMMA	=================
%%	=============================================
\begin{lemma}
\label{lmm:ConstantRate}
In a schedule that ends at $T^{\rm{opt}}$, the rate pair does not change {\emph{within}} epochs in $[0,T^{\rm{opt}}]$. 
\end{lemma}

{\emph{Proof.}} 
 From Corollary~\ref{corollary:ConstantPower}, power level stays constant during epochs in an optimal schedule. Now, suppose the sender changes its rate pair at some point during an epoch, while the power is constant at $P$. Let the lengths of the two slots as $\beta t$ and $(1-\beta)t$ and the rate pairs in the $1^{st}$ and $2^{nd}$ slots as ($r_{11}$, $r_{21}$) and ($r_{12}$, $r_{22}$). Due to the concavity of $h_2(P,r)$ in $r$, setting $r_1$ to the average rate only improves $r_2$
\begin{equation*} 
h_2(P,\beta r_{11} + (1-\beta) r_{12}) \geq \beta h_2(P,r_{11}) + (1-\beta) h_2(P,r_{12}).
\end{equation*} 
Hence, by equating the rate pair, at least the same number of bits can be transmitted at the same time. 
\enp

The next result is an observation of the structure of the basic solution when there is {\emph{only one energy harvest}} (the one at $t=0$).
%%	=============================================
%%	=================	LEMMA	=================
%%	=============================================
\begin{lemma}
\label{lmm:Tmin_OneEpoch}
\begin{emph}
 Suppose $E_i=0~\forall i>1$ in the system model in Fig.\ref{fig:System_Model}. To minimize the overall transmission duration, the sender finishes transmission to both users at the same time.
\end{emph}
\end{lemma}

{\emph{Proof.}} To reach contradiction, suppose that in an optimal solution, the sender finishes transmission to one of the users before the other. This means that the rate pair changes at some point (when the transmission of one of the users ends before the other), although no new energy has been harvested. By Lemma~\ref{lmm:ConstantRate}, averaging the power levels and rates and using one rate pair continuously would enable us to send at least the same number of bits during the same time. This contradicts the optimality of the original solution. \enp

Lemma~\ref{lmm:Tmin_OneEpoch} tells us that the ratio of the rates $r_1/r_2$ is equal to the ratio of the bits $B_1/B_2$. Then, for the AWGN case from \eqref{eq:Bcast_R1} and \eqref{eq:Bcast_R2} the ratio of powers, $\alpha$, can be found by setting:
\begin{equation*}
r_1 = \frac{1}{2}\log_{2}(1+\frac{s_{1}p_{1}}{\sigma^{2}}), r_2 = \frac{1}{2}\log_{2}(1+\frac{s_{2}p_{2}}{s_{2}p_{1}+\sigma^{2}})
\end{equation*}
where $p_1=P\alpha(B_1,B_2),$ $p_2=P(1-\alpha(B_1,B_2))$. Using $r_1/r_2=B_1/B_2$, one can obtain 
\[\left(1+\frac{\alpha s_{1}P}{\sigma^{2}}\right)^{B_2} = \left(1+\frac{(1-\alpha)s_{2}P}{\alpha s_{2}P+\sigma^{2}}\right)^{B_1}.\]
Solving
\iffalse
\begin{eqnarray*}
\frac{B_1}{B_2}=\frac{\frac{1}{2}\log_{2}(1+\frac{\alpha s_{1}P}{\sigma^{2}})}{\frac{1}{2}\log_{2}(1+\frac{(1-\alpha)s_{2}P}{\alpha s_{2}P+\sigma^{2}})}
\end{eqnarray*}
Solving 
\[\left(1+\frac{\alpha s_{1}P}{\sigma^{2}}\right)^{B_2} = \left(1+\frac{(1-\alpha)s_{2}P}{\alpha s_{2}P+\sigma^{2}}\right)^{B_1}\]
\fi
for $\alpha$ and substituting into \eqref{eq:Bcast_R1} 
and \eqref{eq:Bcast_R2} yields a rate pair ($r_1$, $r_2$), for a given value of $P$. 

Before discussing how to find the right value of $P$, it will be illustrative to present an alternative proof for Lemma~\ref{lmm:Tmin_OneEpoch}. Suppose the sender has a power level $P$ to use. The question is to obtain the minimum termination time for all the bits, $T_{\rm min}$, given by the following: 
\begin{eqnarray*}
T_{\rm min}&=&\min(\max(\frac{B_1}{r_1},\frac{B_2}{r_2}))=f(r_1,r_2)\\
&&=\left\{ \begin{array}{rl}
 \frac{B_1}{r_1} &\mbox{ $if$ $\frac{r_1}{r_2}<\frac{B_1}{B_2}$} \\
 \frac{B_2}{r_2} &\mbox{ $if$ $\frac{r_1}{r_2}>\frac{B_1}{B_2}$}
       \end{array} \right.
\end{eqnarray*}

\begin{figure}[htpb]
\centering \includegraphics[scale=0.4]{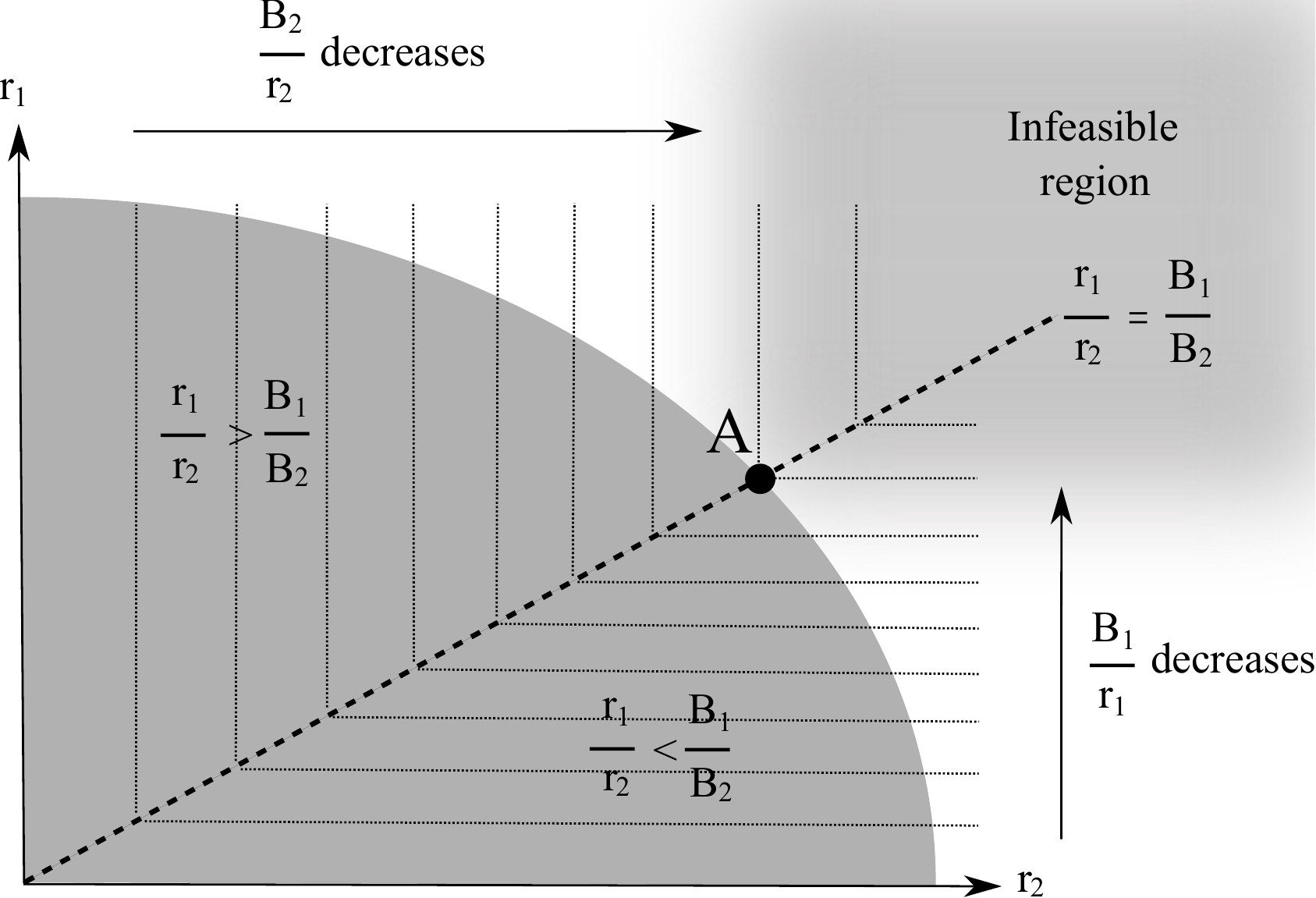}
\caption{Illustration of Lemma~\ref{lmm:Tmin_OneEpoch}: The rate pair ($r_1$, $r_2$) that minimizes overall transmission time for the sender to transmit $B_1$ and $B_2$ bits to each user is at point A.(cf. properties (1)-(4), the shaded rate region is convex.)}
\label{fig:Tmin_ConstantRate}
\end{figure}

The contours of constant $f(r_1,r_2)$ are shown in Figure~\ref{fig:Tmin_ConstantRate}. The value of $f$ gets smaller as we move outward from the rate region. The last contour that still touches the region, touches it at the point labelled A. Therefore, $T_{\rm min}$ is obtained by the rate pair at point A on the boundary of the rate region, which satisfies  $r_1/r_2=B_1/B_2$. 

Given $E$, $B_1$, and $B_2$, determining $T_{\rm min}$ entails solving a nonlinear equation which can be done iteratively (see Algorithm~\ref{alg:Find_Tmin}) using the bisection method. The total energy used to transmit $B_1$ and $B_2$ bits, given by $f(T)=T.g(r_1,r_2)=T.g(B_1/T, B_2/T)$ is convex, monotonically decreasing in $T$. Combining this with our initial assumption about $E$ being large enough to satisfy the minimum energy per bit requirement ($E>\lim_{T\to \infty} T.g(B_1/T, B_2/T)$), there is always a unique smallest value $T$ for which $T.g(r_1,r_2)$ is just below $E$. In the bisection method, the objective is to find the single root of the equation $f(T)-E=0$. Assume that we have an upper bound $T_{\rm upper}>T_{\rm min}$. Starting with the initial domain interval [$T^{\rm min}=0~,~T^{\rm max}=T_{\rm upper}$], at each iteration the domain is bisected and the subinterval in which the root $T_{\rm min}$ lies is selected as follows. If $f(T)-E<0$, then we set $T^{\rm max}=T$, otherwise we set $T^{\rm min}=T$ and $T$ is selected to be the mid-point of the updated interval for the next iteration. Algorithm converges to the unique solution as the domain is continuous and can be terminated within a certain arbitrarily small tolerance $\epsilon>0$ in a practical implementation. This iterative method has been used in generating the numerical examples given later in the paper where its complexity is also discussed.

\begin{algorithm}
\caption{Algorithm to find $T_{\rm min}$ as stated in Lemma~\ref{lmm:Tmin_OneEpoch}}
\label{alg:Find_Tmin}
\begin{algorithmic}[1]
\small
\newcommand{\algorithmicprocedure}{\textbf{procedure}\ }
\newcommand{\algorithmicfunction}{\textbf{function}\ }
\newcommand{\algorithmicprocend}{\textbf{end procedure}\ }
\newcommand{\algorithmicfunctionend}{\textbf{end function}\ }
\STATE \algorithmicprocedure \verb+[+$T$\verb+] = Find_Tmin_One_Epoch(+$E$\verb+,+$B_1$\verb+,+$B_2$\verb+,+$T_{\rm upper}$\verb+)+
\STATE $T^{\rm min} \gets 0$, $T^{\rm max} \gets T_{\rm upper}$, $T \gets T^{\rm max}$, $\tilde{E} \gets 0$
\LOOP
	\STATE $T = (T^{\rm min} + T^{\rm max})/2$ 
	\STATE $\tilde{E} \gets T.g(B_1/T, B_2/T)$\\
	\IF{$(E - \tilde{E}) > \epsilon$}
		\STATE $T^{\rm max} \gets T$ \COMMENT{Less than $E$ units of energy is used for transmission. In order to make $\tilde{E}$ closer to $E$, we need to decrease $T$ while transmitting exactly $B_1$ and $B_2$ bits. Reducing $T^{\rm max}$ guarantees this operation in the next iterate.}
	\ELSIF{$(\tilde{E} - E) > \epsilon$} 
		\STATE $T^{\rm min} \gets T$ \COMMENT{More than $E$ units of energy is used for transmission. In order to make $\tilde{E}$ closer to $E$, we need to increase $T$ while transmitting exactly $B_1$ and $B_2$ bits. Raising $T^{\rm min}$ guarantees this operation in the next iterate.}
	\ELSE
		\RETURN
	\ENDIF
\ENDLOOP
\STATE \algorithmicprocend
\normalsize
\end{algorithmic}
\end{algorithm}

The following extends the result of Lemma~\ref{lmm:Tmin_OneEpoch} to the case with two energy harvests.
%%	=============================================
%%	=================	LEMMA	=================
%%	=============================================
\begin{lemma}
\label{lmm:Tmin_TwoEpoch}
Suppose $E_i=0~\forall i>2$ in the system model in Fig.~\ref{fig:System_Model}. To minimize the overall transmission duration the sender finishes transmission to both users at the same time.
\end{lemma}
\iffalse
\begin{figure}[htpb]
\centering \includegraphics[scale=0.5]{Lemma3_Proof_1.pdf}
\caption{The setting of Lemma~\ref{lmm:Tmin_TwoEpoch}: Given two energy harvests, to minimize the overall transmission duration of $B_1$ and $B_2$ bits to each user, the sender finishes the transmission to both users at the same time, i.e. $T_1=T_2$.}
\label{fig:TwoEpochs}
\end{figure}
\fi
{\emph{Proof.}} 
%\begin{enumerate}
%\item 
The proof will make use of Lemmas~\ref{lmm:ConstantRate} and~\ref{lmm:Tmin_OneEpoch}. Take any rate pair $(r_{11}, r_{21})$ for the $1^{st}$ epoch resulting in $b_{11}=r_{11}\xi_1,~b_{21}=r_{21}\xi_1$ bits being transmitted to the two users. This leaves $b_{12}=B_1-b_{11},~b_{22}=B_2-b_{21}$ bits to be sent in the $2^{nd}$ epoch. By Lemma~\ref{lmm:Tmin_OneEpoch}, total time to finish these remaining bits, which is $T^{\rm{opt}}(2)=T_{\rm min}=\min(\max(T_1,T_2))$ will be minimized by setting $T_1=T_2$. More explicitly,
\begin{eqnarray*}
T_{\rm min}&=&\min(\max(T_1,T_2))=\min(\max(\frac{b_{12}}{r_{12}},\frac{b_{22}}{r_{22}}))\\
&&=\min(\max(\frac{B_1-r_{11}\xi_1}{r_{12}},\frac{B_2-r_{21}\xi_1}{r_{22}}))\\
&&= \left\{ \begin{array}{rl}
\frac{B_1-r_{11}\xi_1}{r_{12}} &\mbox{ $if$ $\frac{r_{12}}{r_{22}}<\frac{B_1-r_{11}\xi_1}{B_2-r_{21}\xi_1}$} \\
\frac{B_2-r_{21}\xi_1}{r_{22}} &\mbox{ $if$ $\frac{r_{12}}{r_{22}}>\frac{B_1-r_{11}\xi_1}{B_2-r_{21}\xi_1}$}
       \end{array} \right.
\end{eqnarray*}

\begin{figure}[htpb]
\centering \includegraphics[scale=0.4]{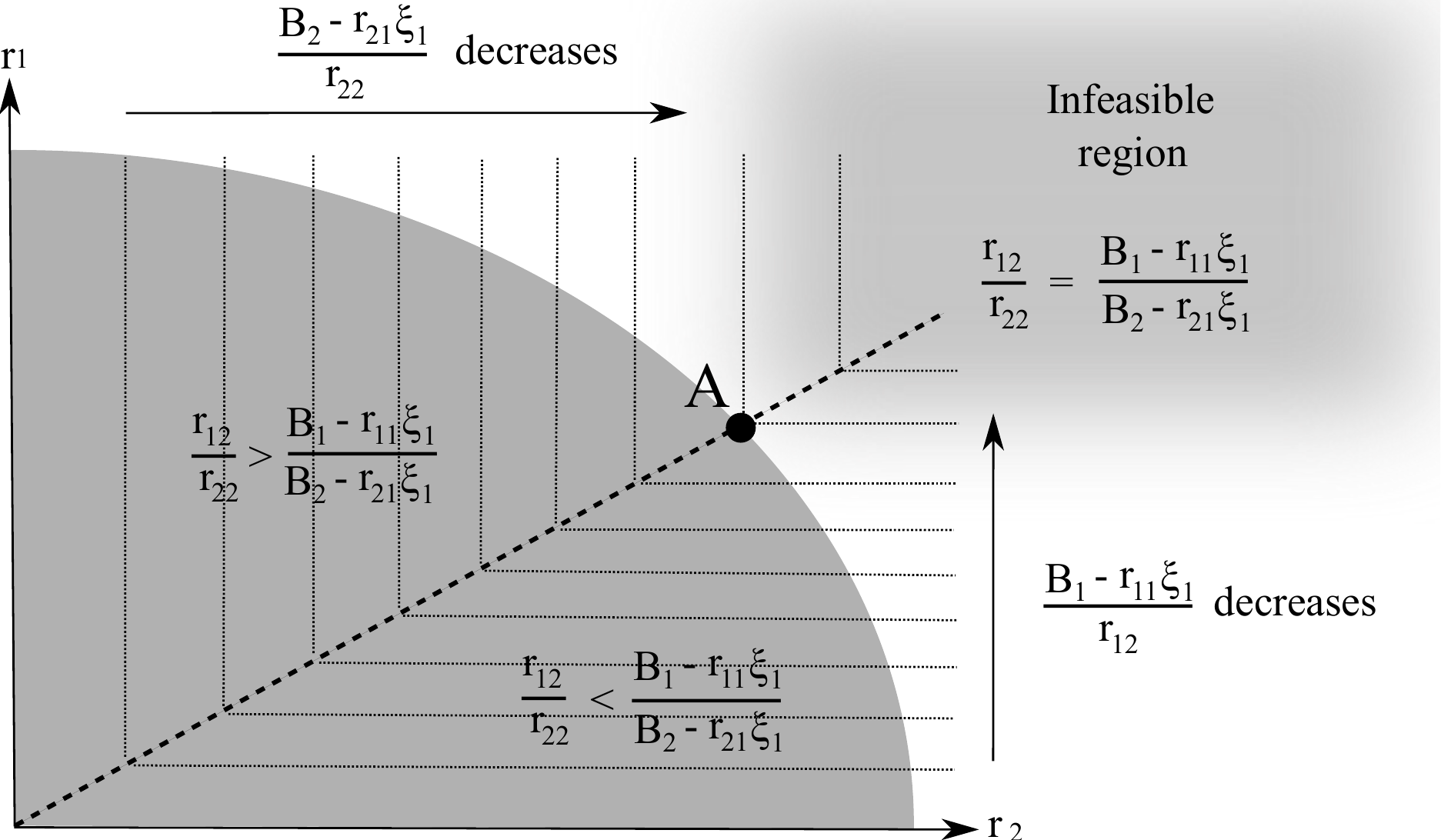}
\caption{Illustration of the proof of Lemma~\ref{lmm:Tmin_TwoEpoch}.}
\label{fig:Tmin_TwoEpoch}
\end{figure}

$T^{\rm{opt}}(2)$ is obtained by the rate pair satisfying $r_{12}/r_{22}=(B_1-r_{11}\xi_1)/(B_2-r_{21}\xi_1)$ as point A on achievable rate region shown in Fig.~\ref{fig:Tmin_TwoEpoch}. By Lemma~\ref{lmm:ConstantRate}, for the resulting two-epoch schedule to be optimal, a constant power and rate pair must have been used in the $1^{st}$ epoch, and we have just proved that $T_1=T_2$ for any constant choice in the $1^{st}$ epoch.\enp

%%	=============================================
%%	=================	LEMMA	=================
%%	=============================================
Finally, we generalize the first result of Lemma~\ref{lmm:Tmin_TwoEpoch}, to a general number of energy harvests.
\begin{lemma}
\label{lmm:Tmin_GeneralEpoch}
\begin{emph} 
Consider the system model with an arbitrary number of energy harvests described in Fig. \ref{fig:System_Model}. In a schedule that achieves $T^{\rm{opt}}$, the transmission to both users ends at the same time. 
\end{emph}
\end{lemma}

{\emph{Proof.}} The claim has been proved for $k=2$ energy harvests, in Lemma~\ref{lmm:Tmin_TwoEpoch}. We will prove the general case by induction. Suppose that there are $k$ energy harvests with the $k^{th}$ one at time $t_{k}$, and the induction hypothesis holds, such that the optimal scheduler finishes transmission to both users at $T^{\rm{opt}}(k)$. Now, consider adding a new energy harvest at time $t_{k+1}$. We have the following possible cases:
\begin{enumerate}
\item $t_{k+1}\geq T^{\rm{opt}}(k)$: By the time the $(k+1)^{st}$ energy harvest arrives, the transmission has been completed, so by causality this energy harvest cannot help, and $T^{\rm{opt}}(k+1)=T^{\rm{opt}}(k)$.
\item $t_{k+1} < T^{\rm{opt}}(k)$: In this case, the $(k+1)^{st}$ harvest will be used, to reduce the completion time. Starting at time $t_k$, the sequence of rate pairs will change from $\{(r_{1i},r_{2i})\}$ for $i=1,\ldots,k$ to $\{(\tilde{r}_{1i},\tilde{r}_{2i})\}$ for $i=1,\ldots,k$ and $(\tilde{r}_{1(k+1)},\tilde{r}_{2(k+1)})$ for the newly added epoch. As in the proof of Lemma~\ref{lmm:Tmin_TwoEpoch}, whatever the number of bits allocated to the new epoch is, this rate pair will have a slope equal to the ratio of the number of bits remaining for this epoch. Hence, the bits will be terminated at some time $T^{\rm{opt}}(k+1)\leq T^{\rm{opt}}(k)$. \enp
\end{enumerate}

We are now ready to state the broadcast transmission scheduling problem as an optimization problem. From Section~\ref{sec:bc}, for a given rate pair, there corresponds a unique power level $P$ given by $g(r_1,r_2)$ such that this rate pair is on the boundary of the rate region with power constraint $P$. The function $g(r_1,r_2)$ is strictly convex and continuously differentiable in $r_1$ and $r_2$. Using Lemmas~\ref{lmm:ConstantRate} and~\ref{lmm:Tmin_GeneralEpoch}, the problem can be written in terms of epoch rates. 

Given $B_1, B_2$, and the sequence $\{E_i\}$, supposing the problem is feasible (the total amount of energy is sufficient for transmitting the total number of bits), one can find an upperbound for the transmission completion time, $T^{\rm up}$ in several ways. A simple one (which is possible when $E_1$ is sufficient for transmitting the total number of bits), is to set the power so low such that only the first harvest is used to transmit all the bits. A much better upperbound will be obtained by the procedure that will be described within the initialization step of the FlowRight algorithm, in Section~\ref{sec:solution}.

Given an upperbound for completion time, $T^{\rm up}$, we set $k^{\rm up}$ equal to the index of the last energy harvest before this time, that is, $k^{\rm up} = \max \{i:\sum_{j=1}^{i} \xi_j \leq T^{\rm up}\}$. An optimal solution will use at most $k^{\rm up}$ harvests, and WLOG, remaining harvests can be ignored. Hence the problem reduces to finding $T^{\rm opt}(k^{\rm up})$:
 
\begin{problem}
\label{pr:MultiuserScheduling}
\noindent{\bf Transmission Time Minimization of Data Available at the Beginning on an Energy Harvesting Broadcast Channel:}
\small \begin{align}
\noindent \mbox{Minimize:  } &T=T(\{ (r_{1i},r_{2i})\}_{1 \leq i \leq k^{\rm up}}) \nonumber \\
\noindent \mbox{subject to: } 
&r_{1i},r_{2i}\geq 0,\;\; 1 \leq i \leq k^{\rm up} \nonumber\\
&0<T \leq T^{\rm up} \nonumber\\
&\sum_{i=1}^{k} g(r_{1i},r_{2i})\xi_{i} \leq \sum_{i=1}^{k} E_i \label{eq:OptConst1}\\
&for \mbox{ }k=1,2,...,k^{*}=\max\{i:\sum_{j=1}^{i} \xi_j \leq T\}\nonumber\\
&\sum_{i=1}^{k^{*}} g(r_{1i},r_{2i})\xi_{i} + 
	 g\left(\frac{(B_1-\sum_{i=1}^{k^{*}}r_{1i}\xi_i)^+}{\left(T-\sum_{i=1}^{k^{*}} \xi_{i}\right)},      		    \frac{(B_2-\sum_{i=1}^{k^{*}}r_{2i}\xi_i)^+}{\left(T-\sum_{i=1}^{k^{*}} \xi_{i}\right)}\right)\left(T-\sum_{i=1}^{k} \xi_{i}\right)=\sum_{i=1}^{k+1} E_i \label{eq:OptConst2}
\end{align}\normalsize
\end{problem}

The set of constraints in \eqref{eq:OptConst1} ensure that energy causality is respected. At any time during transmission, the sender should have consumed {\emph{at most}} the energy harvested up to that point, whereas by the end of transmission, it should have consumed {\emph{all}} the harvested energies up to that instant. The constraint in \eqref{eq:OptConst2} ensures that all the bits of each user have been transmitted by the time $T$. Note that, by assigning nonzero values to all $k^{\rm up}$ rates, one obtains a continuum of values of $T$ that satisfy the constraints (note the $(\;\;)^+$ used in the last constraint which sets the result to zero whenever the argument is negative), but the infimum of these, $T^{\rm{opt}}$, is the solution of the problem. We shall define $n^{\rm{opt}}=\min\{i:\sum_{j=1}^{i} \xi_j \geq T^{\rm{opt}}\}$, \ie, the index of the last harvest used by a solution that achieves $T^{\rm{opt}}$ (It can easily be shown any such solution completely consumes all harvests from $i$ to $n^{\rm{opt}}$.)

This is not a standard convex optimization problem due to the objective appearing in the final equality constraint.  Yet, we will establish that this minimization problem can be solved iteratively using an adaptation of the FlowRight algorithm~\cite{Eu04}\iffalse \footnote{Flowright is essentially nonlinear block coordinate descent method (a.k.a. Gauss-Seidel method.) The convergence of Gauss-Seidel iterations does not require a convex objective, only that the optimization with respect to a subset of variables (in our case rates of two epochs) leads to a unique minimum in terms of the global objective~\cite{Bert95}. Yet, convergence and optimality proofs of FlowRight will NOT rely on this observation or on convexity of $T$ in the rate vector, while we do suspect that $T$ is convex in the rate vector.}\fi. Before moving on to the solution, we present our final observations in the optimal schedule in Theorems~\ref{thm:OptPowerAllocation}-\ref{thm:OptRateAllocation}~\footnote{Similar claims to those listed in Theorem~\ref{thm:OptRateAllocation} have been proven in \cite{YYOOSO2010} through the observation that there is a cut-off level for the total power, below which the weaker user is assigned zero rate.} using the general result presented in Lemma~\ref{lmm:CloserRate}.

%%	=============================================
%%	=================	THEOREM	=================
%%	=============================================
\begin{theorem}
\label{thm:OptPowerAllocation}
\emph{In an optimal schedule,}
\begin{enumerate}
\item \emph{Powers assigned to epochs are monotonically nondecreasing, \ie, $P_1 \leq P_2 \leq...\leq P_{n^{\rm opt}}$.}
\item \emph{Energy consumed in any constant power band equals the total energy harvested within that band.}
\item \emph{The power assignment to epochs, $\textbf{P}^{\rm opt}=[P_1, P_2, ..., P_{n^{\rm opt}}]$, is unique.}
\end{enumerate}
\end{theorem}

{\emph{Proof.}} 
\begin{enumerate}
\item Suppose in an optimal solution we can find $i$ s.t. $P_i > P_{i+1}$. From Lemma~\ref{lmm:CloserPowers}, equalizing power over these epochs (this never violates causality as it corresponds to deferring the use of energy), one could find a rate pair with which more bits can be transmitted to each user. This contradicts the optimality of the original solution.

\item Suppose that $P_i=P_s\neq P_{s+1},~s-m \leq i \leq s < n_{\rm{opt}}$ for some band of length $m<s$ such that $\sum_{i=s-m}^{s} E_i^{\rm{opt}} < \sum_{i=s-m}^{s} E_i$. But by Part-1, $P_{s+1}>P_s$ so a positive amount of energy (up to $\sum_{i=s-m}^{s} E_i-\sum_{i=s-m}^{s} E_i^{\rm{opt}}$ units) can be transferred from epoch $s+1$ to $s$ to equalize their powers, which by  Lemma~\ref{lmm:CloserPowers}, allows rate pairs that send at least the same number of bits to each user. Keeping the rates and powers in the rest the same, the overall improved schedule is obtained, which contradicts the optimality of the original allocation.

\item Suppose that there are two different optimal power allocation vectors, $\textbf{P}^A$ and $\textbf{P}^B$, where $P^A_k=P^B_k$ for $k={1,2,..,i-1}$ and $P^A_i < P^B_i$. From Part-1, power levels are  monotonically nondecreasing
in the optimal schedule. In this case, if $P^A_k$ for $k \geq i$ stays constant, we have $\sum_{k=i+1}^{n} P^A_k \xi_k < \sum_{k=i+1}^{n} P^B_k \xi_k$, else $\exists~j:\{P^A_i < P^A_{i+j},  1 \leq j \leq n-i\}$ and we have $\sum_{k=i+1}^{j-1} P^A_k \xi_k < \sum_{k=i+1}^{j-1} P^B_k \xi_k$, both contradicting Part-2. \enp
\end{enumerate}

%%	=============================================
%%	=================	LEMMA	=================
%%	=============================================
\begin{lemma}
\label{lmm:CloserRate}
Suppose the sender uses different rates for the \emph{stronger} user in the intervals $(\tau_1, \tau^*)$, $(\tau^*, \tau_2)$, such that $\tau_1<\tau^*< \tau_2$. Keeping powers levels and the number of bits transmitted to the \emph{stronger} user in $(\tau_1, \tau_2)$ constant, a larger number of bits can be sent to the \emph{weaker} user in $(\tau_1, \tau_2)$ by bringing the rates of the \emph{stronger} user closer to each other if feasible.
\end{lemma}

{\emph{Proof.}} Consider the notation in Figure~\ref{fig:CloserPowers} and the case $r_{11} < r_{12}$. Keeping the avg. rate of user 1, $\bar{r_{1}}$, constant, set $r_{11}$ to $r_{11}^{'}$, $r_{12}$ to $r_{12}^{'}$ s.t. $r_{11} \leq r_{11}^{'} \leq r_{12}^{'} \leq r_{12}$ by transferring a certain amount of bits belonging to \emph{stronger} user are transferred from the $2^{nd}$ slot to the $1^{st}$. This is feasible unless $r_{11}$ is already maximal for the given power level (i.e. $r_{21}=0$).
\iffalse
\begin{equation}
r_{11}^{'}= r_{11} + (1-\beta)\Delta r \mbox{~,~}r_{12}^{'}= r_{12} - \beta \Delta r.
\label{eq:Del_r}
\end{equation}
\fi
Avg. rate of $2^{nd}$ user over the whole duration is increased from its original level, $\bar{r_{2}}$, to:
\begin{eqnarray}
\bar{\bar{r_{2}}}&=& h_2(P_1,r_{11}^{'})\beta + h_2(P_2,r_{12}^{'})(1-\beta) \nonumber \\
&> & h_2(P_1,r_{11})\beta + h_2(P_2,r_{12})(1-\beta)=\bar{r_{2}} \label{eq:Avg_2nd_User_Rate_2}
\end{eqnarray}
\eqref{eq:Avg_2nd_User_Rate_2} follows from the fact that
\begin{equation}
 h_2(P_1,r_{11}^{'})\beta + h_2(P_2,r_{12}^{'})(1-\beta) - h_2(P_1,r_{11})\beta - h_2(P_2,r_{12})(1-\beta) \geq 0
\label{eq:rate_f}
\end{equation}
for all $\beta\neq \{0,1\}$ (with equality achieved at $\beta=0,1$), unless $r_{21}=0$, as proved in App.~\ref{subsec:Appendix_closer_rates}
In the remaining case, $r_{11}>r_{12}$, set $r_{12} \leq r_{12}^{'} \leq r_{11}^{'} \leq r_{11}$, which is feasible unless unless $r_{22}=0$, and strictly improves the avg. rate for user 2. \enp

\iffalse
We conclude that keeping $\bar{r_1}$ constant, one can find rate pairs such that at least the same number of bits can be transmitted to the users within the same duration by reallocating rates of \emph{stronger} user.
\fi

%%	=============================================
%%	=================	THEOREM	=================
%%	=============================================
\begin{theorem}
\label{thm:OptRateAllocation}
\emph{In an optimal schedule,}
\begin{enumerate}
\item \emph{the stronger user's rate is monotone nondecreasing, \ie, $r_{11} \leq r_{12} \leq...\leq r_{1n^{\rm{opt}}}$;}
\item \emph{if $r_{1(i+1)} \neq r_{1i}$ for some $0<i<n^{\rm opt}$, then $r_{2i}=0$, \ie, if the stronger user's rate changes at the start of the $(i+1)^{th}$ epoch, the weaker user's rate was zero during the $i^{th}$ epoch;}
\item \emph{the weaker user's rate is monotone nondecreasing, \ie, $r_{21} \leq r_{22} \leq...\leq r_{2n^{\rm{opt}}}$;}
\item \emph{The vector of rate pairs, $\textbf{R}^{\rm{opt}}=[(r^{\rm opt}_{11},r^{\rm opt}_{21}), ..., (r^{\rm opt }_{1n^{\rm opt }},r^{\rm opt}_{2n^{\rm opt }})]$, is unique.}
\end{enumerate}
\end{theorem}

{\emph{Proof.}}
\begin{enumerate}
\item Suppose the rate of the \emph{stronger} user decreases at some point, \ie,  $r_{1i}>r_{1(i+1)}$ for some $i$. From Part-1 of Theorem~\ref{thm:OptPowerAllocation} and Lemma~\ref{lmm:CloserRate}, at least the same number of bits can be sent to each user (and more to at least one) in epochs $(i,i+1)$ by assigning the strong user the average rate $\bar{r_1}$. 

\item Suppose that in an optimal schedule the weaker user's rate changes at the $(i+1)^{th}$ epoch and $r_{2i}\neq 0$. If the rate of the \emph{stronger} user changes, it can only increase, \ie, $r_{1i}<r_{1(i+1)}$, by Part-1. From Lemma~\ref{lmm:CloserRate}, the schedule could only be improved by bringing $r_{1i}$ and $r_{1(i+1)}$ closer to each other using the energy available for the \emph{weaker} user at the $i^{th}$ epoch, if possible. Hence, the only reason why the rate of the \emph{stronger} user can increase at the $(i+1)^{th}$ epoch is that there is no feasible energy available to equalize $r_{1i}$ and $r_{1(i+1)}$, which contradicts $r_{2i}\neq0$.
\item Suppose that in an optimal schedule $r_{2i}>r_{2(i+1)}$. From Part-2, $r_{1i} = r_{1(i+1)} = \bar{r_1}$ if $r_{2i} \neq 0$. By Part-1 of Theorem~\ref{thm:OptPowerAllocation} and $2^{nd}$ property of the rate region, $r_{2i} = h_2(P_i,\bar{r_1}) \leq h_2(P_{i+1},\bar{r_1}) =r_{2(i+1)} $ which contradicts initial rate assumption.
\item To reach contradiction, suppose that there are two distinct optimal rate-pair vectors, $\textbf{R}^A$ and $\textbf{R}^B$, where $(r^A_{1k}, r^A_{2k})=(r^B_{1k}, r^B_{2k})$ for $k={1,2,..,i-1}$ and $r^A_{1i} < r^B_{1i}$. Using Part-3 of Theorem~\ref{thm:OptPowerAllocation}, we have $r^A_{2i}=h_2(P_i,r^A_{1i})>h_2(P_i,r^B_{1i}) \geq 0$. From Part-2, $r^A_{1j}=r^A_{1i} < r^B_{1i} \leq r^B_{1j}~\forall~j \in \{i+1,...,n\}$. Hence, fewer bits will be transmitted by $\textbf{R}^A$ than $\textbf{R}^B$, which contradicts the optimality of $\textbf{R}^A$.\enp
\end{enumerate}

From Theorems~\ref{thm:OptPowerAllocation} and \ref{thm:OptRateAllocation}, we conclude that \emph{the optimal schedule is unique} (henceforth abbreviated as OPT.) The next section is devoted to the solution of Problem \ref{pr:MultiuserScheduling}.

\section{Optimal Offline Scheduling with the FlowRight Algorithm}
\label{sec:solution}
\iffalse
FlowRight is an iterative algorithm proposed in the earlier literature~\cite{Eu04}\iffalse to solve minimum-energy scheduling problems over multiple-access channels, broadcast channels, and channels with fading when packets of all users need to be transmitted before a deadline T.\fi Through a number of steps, it is adapted here to solve Problem~\ref{pr:MultiuserScheduling}. 
\fi
FlowRight stars from a feasible initial schedule, and progresses iteratively. Each iteration strictly improves the schedule (decreases $T$), which ultimately converges to the unique optimal $T^{\rm{opt}}$.

\textbf{Initialization:} The energy consumed in each epoch is set precisely equal to the energy harvested at the beginning of that epoch. This schedule is feasible and $P_i^0=E_i/\xi_{i}$. Given $P_i^0,~i \in \{1,2,...\}$, one can assign rate pairs $(r_{1i}^0,r_{2i}^0)~i \in \{1,2,...\}$ on the achievable rate region boundary such that $r_{1i}^0/r_{2i}^0 =B_{1}/B_{2}$. Let $n^{\rm up}=\underset{i}{\operatorname{argmin}}\left\{r_{1i}^0=0,r_{2i}^0=0\right\}$. Algorithm~\ref{alg:Init} presents a pseudo-code for this initialization.

\begin{algorithm}
\caption{Initialization of FlowRight}
\label{alg:Init}
\begin{algorithmic}[1]
\small
\STATE $i \gets 0$
\WHILE{$B_1 \neq 0 \mbox{~} || \mbox{~} B_2 \neq 0$} 
	\STATE i++
	\STATE Select $(r_{1i}^{0}, r_{2i}^{0})$ such that: $g(r_{1i}^{0}, r_{2i}^{0}) \gets E_{i}/\xi_i$ and $r_{1i}^{0}/r_{2i}^{0} \gets B_{1}/B_{2}$
	\STATE $B_1 \gets B_1-r_{1i}^{0}\xi_i$ \COMMENT{Update remaining bits of $1^{st}$ user.}
	\STATE $B_2 \gets B_2-r_{2i}^{0}\xi_i$ \COMMENT{Update remaining bits of $2^{nd}$ user.}		
\ENDWHILE
\STATE $n^{\rm up} \gets i$ \COMMENT{Set the initial number of epochs to be considered.}
\normalsize
\end{algorithmic}
\end{algorithm}
After initialization, FlowRight performs \textit{local optimizations} on pairs of epochs sequentially, \ie, on epochs $(1,2)$, $(2,3)$, $(3,4)$, ... , until all epoch pairs are processed. This completes one iteration of the algorithm. Then, it continues with the next iteration, again performing local optimization on pairs of epochs at a time. The algorithm terminates after $K$ iterations such that $K=\min\left\{k:(T^{k-1}-T^{k})<\epsilon\right\}$, where $T^{k}$ is the transmission completion time at the end of $k^{th}$ iteration. 

\textbf{Local optimization:} Let $E_{i}^{k}$ be the energy consumed at the $i^{th}$ epoch and $n_{k}\leq n^{\rm up}$ is the number of epochs used at the end of the $k^{th}$ iteration. Then, $E_i^{0}=E_i$, $i=1,2,...,n^{\rm up}$. Also, let $b_{ji}^{k}$ be the number of bits transmitted to $j^{th}$ user at $i^{th}$ epoch at the end of $k^{th}$ iteration. Now, consider the epoch pair $(i,i+1)$, $i \in \{1,2,..,n_{k}-1\}$. Local optimization aims to transmit the total $b_{j}^{k}=b_{ji}^{k-1}+b_{j(i+1)}^{k-1}$ bits in the minimum amount of time by using the total available energy, \ie, $E_{(i,i+1)}^{k}=E_{i}^{k-1} + E_{i+1}^{k-1}$ while \textbf{respecting energy causality}, \ie, using at most $\sum_{m=1}^{i} E_m^{0}-\sum_{m=1}^{i-1} E_m^{k}$ amount of energy in the $i^{th}$ epoch (see the term $E_{max_i}^k$ in Algorithm~\ref{alg:FlowRight}). After a local optimization, we update the rate of the $j^{th}$ user in the $i^{th}$ epoch to $r_{ji}^{k}$. These are the final values of the $i^{th}$ epoch rates as of the end of the $k^{th}$ iteration. We then reset the rate of the $j^{th}$ user in the $(i+1)^{th}$ epoch to $r_{j(i+1)}^{k-1}$, unless $(i+1) = n_k$. The $(i+1)^{th}$ epoch rates as of the $k^{th}$ iteration are finalized after the local optimization on the epoch pair $(i+1,i+2)$ has been performed.

Then, we continue with local optimization on the epoch pair $(i+1,i+2)$. We proceed in this way to obtain $(r_{1i}^{k},r_{2i}^{k})$ for $i=1,2,...,n_{k}$. When the $k^{th}$ iteration is finished, we start from the beginning and update rates two epochs at a time similar to the above. It will be shown in Theorem~\ref{thm:flowright_stops} that transmission completion time strictly decreases after each iteration, and the number of epochs used, $n_{k}$, is non-increasing from iteration to iteration. We terminate after $K$ iterations, where $K=\min\left\{k:(T^{k-1}-T^{k})<\epsilon\right\}$. Algorithm~\ref{alg:FlowRight} is the main pseudo-code for FlowRight.

\begin{algorithm}
\caption{FlowRight}
\label{alg:FlowRight}
\begin{algorithmic}[1]
\small
\STATE {k $\gets$ 0, $T^{0} \gets T^{up}$}
\REPEAT 
	\STATE {k++} 
	\FOR{$i = 1$ \TO $(n_{k-1}-1)$} 
		\STATE $b_1^{k} \gets b_{1i}^{k-1} +  b_{1(i+1)}^{k-1}$ 
		\STATE $b_2^{k} \gets b_{2i}^{k-1} +  b_{2(i+1)}^{k-1}$
		\STATE $E_{\rm max_i}^{k} \gets \min\left\{(\sum_{m=1}^{i} E_m^{0}-\sum_{m=1}^{i-1} E_m^{k})~,~(E_i^{k-1}+E_{(i+1)}^{k-1})\right\}$			
		\STATE {[$b_{1i}^{k}$\verb+,+$b_{1(i+1)}^{k-1}$\verb+,+$b_{2i}^{k}$\verb+,+$b_{2(i+1)}^{k-1}$\verb+,+$E_i^k$\verb+,+$E_{(i+1)}^{k-1}$] = \verb+Find_Local_Optimal(+$E_{\rm max_i}^{k}$\verb+,+$E_i^{k-1}$\verb+,+$E_{i+1}^{k-1}$\verb+,+$b_1^k$\verb+,+$b_2^k$\verb+)+} \COMMENT{Perform \textit{Local Optimization} using \textbf{Algorithm~\ref{alg:LocalOpt}}.}
	\ENDFOR
		\IF {$b_{1(n_k-1)}^k$ == 0 \&\& $b_{2(n_k-1)}^k$ == 0}
			\STATE $n_k$=$n_{k-1}-1$ \COMMENT{Reduce the number of epochs for the next iteration.}
		\ELSE
			\STATE $n_k$=$n_{k-1}$ 
		\ENDIF
		\STATE \verb+Calculate_T(&+$T^k$ \verb+)+ \COMMENT{Calculate current transmission completion time after one iteration over the epochs.}

\UNTIL {$(T^{k-1}-T^k)<\epsilon $}
\normalsize
\end{algorithmic}
\end{algorithm}

\textbf{Details of the Local Optimization}
The local iteration step is described in Algorithm~\ref{alg:LocalOpt} which mainly checks whether it is possible to transmit $b_1^k$ and $b_2^k$ bits, \textbf{(by respecting energy causality)} in the minimum amount of time via equalizing power and rates in two epochs (it is optimal from Lemmas~\ref{lmm:CloserPowers} and ~\ref{lmm:CloserRate}), or not. If feasible, then min. transmission time is found by treating the epoch pair $(i,i+1)$ as a single epoch with duration $T_{\rm upper}=\xi_i+\xi_{i+1}$. Then, Alg.~\ref{alg:Find_Tmin} starting with the initial $[T^{\rm min}=0, T^{\rm max}=T_{\rm upper}]$ domain will find the unique solution of local optimization. If power cannot be equalized, then we are at an energy causality boundary. In this case, Alg.~\ref{alg:LocalOpt_Causality}, which can be implemented using the bisection method, is used to find the updated rate pairs $(r_{1i}^k,r_{2i}^k)$. The local optimization given in Alg.~\ref{alg:LocalOpt_Causality} exploits the structure of a locally optimized epoch pair (see Theorem~\ref{thm:flowright_properties} to work efficiently. The proof of this structure is omitted for brevity but follows similar arguments on bisection method of Alg.~\ref{alg:Find_Tmin}.

\begin{algorithm}
\caption{Algorithm to perform Local Optimization on the epoch pair $(i,i+1)$ }
\label{alg:LocalOpt}
\begin{algorithmic}[1]
\newcommand{\algorithmicprocedure}{\textbf{procedure}\ }
\newcommand{\algorithmicfunction}{\textbf{function}\ }
\newcommand{\algorithmicprocend}{\textbf{end procedure}\ }
\newcommand{\algorithmicfunctionend}{\textbf{end function}\ }
\small
\STATE \algorithmicprocedure [$b_{1i}^{k}$\verb+,+$b_{1(i+1)}^{k-1}$\verb+,+$b_{2i}^{k}$\verb+,+$b_{2(i+1)}^{k-1}$\verb+,+$E_i^k$\verb+,+$E_{(i+1)}^{k-1}$] = \verb+Find_Local_Optimal(+$E_{\rm max_i}^{k}$\verb+,+$E_i^{k-1}$\verb+,+$E_{i+1}^{k-1}$\verb+,+$b_1^k$\verb+,+$b_2^k$\verb+,+$\xi_i$\verb+,+$\xi_{i+1}$\verb+)+
\STATE $E_{(i,i+1)}^{k}=E_{i}^{k-1} + E_{i+1}^{k-1}$\\
\STATE $T_{\rm min}$\verb+= Find_Tmin_One_Epoch(+$E_{\rm max_i}^{k}$\verb+,+$b_{1}^{k}$\verb+,+$b_{2}^{k}$\verb+,+$\xi_i+\xi_{i+1}$\verb+)+ \COMMENT{Calculate $T_{\rm min}$ to transmit $b_{1}^{k}$ and $b_{2}^{k}$ bits \textbf{using all the feasible energy for the $i^{th}$ epoch, \ie, $E_{\rm max_i}^{k}$ at $k^{th}$ iteration.}}
\IF {$T_{\rm min} < \xi_i$} 
		\STATE \COMMENT{$E_{\rm max_{i}}^{k}$ is sufficient to transmit all the $b_{1}^{k}$ and $b_{2}^{k}$ bits.}
        \STATE $b_{1i}^{k} \gets b_{1}^{k} \mbox{~,~} b_{2i}^{k} \gets b_{2}^{k} \mbox{~,~} b_{1(i+1)}^{k-1} \gets 0 \mbox{~,~} b_{2(i+1)}^{k-1} \gets 0$ \COMMENT{Update/Reset bits in the $i^{th}$/$(i+1)^{th}$ epoch.}
        \STATE $E_{i}^{k} \gets g(r_{1i}^{k},r_{2i}^{k}).T_{\rm min} \mbox{~,~} E_{i+1}^{k-1} \gets E_{(i,i+1)}^{k}-E_{i}^{k}$ \COMMENT{Update/Reset energy in the $i^{th}$/$(i+1)^{th}$ epoch.}

\ELSE
        \STATE $T_{\rm min}$\verb+= Find_Tmin_One_Epoch(+$E_{(i,i+1)}^{k}$\verb+,+$b_{1}^{k}$\verb+,+$b_{2}^{k}$\verb+,+$\xi_i+\xi_{i+1}$\verb+)+ 
        \STATE $E_i^k \gets g(b_{1}^{k}\xi_i/T_{\rm min}, b_{2}^{k}\xi_i/T_{\rm min})$ \{Update energy within the $i^{th}$ epoch\}
        \IF {$E_{i}^{k} \leq E_{\rm max_i}^{k}$}
        		\STATE \COMMENT{We can equalize powers without violating energy causality} 
                \STATE $b_{1i}^{k} \gets b_{1}^{k}\xi_i/T_{\rm min} \mbox{~,~} b_{2i}^{k} \gets b_{2}^{k}\xi_i/T_{\rm min}$ \COMMENT{Update bits in the $1^{st}$ epoch.}
                \STATE $b_{1(i+1)}^{k-1} \gets b_{1}^{k}-b_{1i}^{k} \mbox{~,~} b_{2(i+1)}^{k-1} \gets b_{2}^{k}-b_{2i}^{k}$ \COMMENT{Reset bits in the $(i+1)^{th}$ epoch.}
                \STATE $E_{i+1}^{k-1} \gets E_{(i,i+1)}^{k}-E_{i}^{k}$ \COMMENT{Reset energy in the $(i+1)^{th}$ epoch.}                
        \ELSE
        		\STATE \COMMENT{We can NOT equalize powers without violating energy causality. Maximum feasible energy should be allocated for the $i^{th}$ epoch.}
        		\STATE $E_{i}^{k} \gets E_{\rm max_i}^{k} \mbox{~,~} E_{i+1}^{k-1} \gets E_{(i,i+1)}^k-E_{\rm max_i}^{k}$ \COMMENT{Update/Reset energy in the $i^{th}/(i+1)^{th}$ epoch.} 	
        		\STATE \verb+[+$r_{1i}^k$\verb+,+$r_{2i}^k$\verb+]=Find_Tmin_Two_Epoch(+$E_i^k$\verb+,+$E_{i+1}^{k-1}$\verb+,+$b_1^k$\verb+,+$b_2^k$\verb+,+$\xi_i$\verb+,+$\xi_{i+1}$\verb+)+\textbf{ (see Algorithm~\ref{alg:LocalOpt_Causality})}.	   				
        		\STATE $b_{1i}^{k} \gets r_{1i}^k \xi_i \mbox{~,~} b_{1(i+1)}^{k-1} \gets  b_1^k-b_{1i}^k \mbox{~,~} b_{2i}^{k} \gets r_{2i}^k \xi_i \mbox{~,~} b_{2(i+1)}^{k-1} \gets b_2^k-b_{2i}^k$ \COMMENT{Update/Reset bits in the $i^{th}$/$(i+1)^{th}$ epoch.}
        \ENDIF
\ENDIF 
\STATE \algorithmicprocend
\normalsize
\end{algorithmic}
\end{algorithm}

\begin{algorithm}
\caption{Algorithm performing \textit{Local Optimization} on the epoch pair $(i,i+1)$ when energy causality is met in the $i^{th}$ epoch. It is based on the observations in Theorem~\ref{thm:flowright_properties}. }
\label{alg:LocalOpt_Causality}
\begin{algorithmic}[1]
\small
\newcommand{\algorithmicprocedure}{\textbf{procedure}\ }
\newcommand{\algorithmicfunction}{\textbf{function}\ }
\newcommand{\algorithmicprocend}{\textbf{end procedure}\ }
\newcommand{\algorithmicfunctionend}{\textbf{end function}\ }
\STATE \algorithmicprocedure \verb+[+$r_{1i}$\verb+,+$r_{2i}$\verb+] = Find_Tmin_Two_Epoch(+$E_i$\verb+,+$E_{i+1}$\verb+,+$b_1$\verb+,+$b_2$\verb+,+$\xi_i$\verb+,+$\xi_{i+1}$\verb+)+
\STATE $r_{1i}^{\rm min} \gets 0$, $r_{1i}^{\rm max} \gets \min\{h_1(E_i/\xi_i, 0)~,~B_1/\xi_i\}$
\STATE $r_{1i} \gets r_{1i}^{\rm max}$, $r_{1(i+1)} \gets 0$, $r_{2i} \gets 0$, $r_{2(i+1)} \gets 0$

\COMMENT{Check whether locally optimal rates of 1st user changes or not.}
\IF {$h_1(E_i/\xi_i, 0)~<~B_1/\xi_i$}
	\STATE $r_{1(i+1)} \gets r_{1i}$\\
	\STATE $T \gets (b_1 - r_{1i} \xi_i)/r_{1(i+1)}$\\	
	\STATE $r_{2i} \gets h_2(E_i/\xi_i,r_{1i})$, $r_{2(i+1)} \gets h_2(E_{i+1}/T,r_{1(i+1)})$\\
	\STATE $\tilde{b}_2=r_{2i} \xi_i + r_{2(i+1)} T$
	\IF{$(\tilde{b}_2 - b_2) > \epsilon$} 	
		\STATE \verb+Find_Tmin_One_Epoch(+$E_{i+1}$\verb+,+$b_1-r_{1i}\xi_i$\verb+,+$b_2-r_{2i}\xi_i$\verb+,+$\xi_{i+1}$\verb+)+ \COMMENT{Calculate minimum transmission time to transmit remaining bits within the $(i+1)^{th}$ epoch using $E_{i+1}$ amount of energy}
		\RETURN
	\ENDIF
\ENDIF

\COMMENT{Locally optimal rates of 1st user are the same. Start Bisection Method.}
\LOOP
	\STATE $r_{1i} = (r_{1i}^{\rm min} + r_{1i}^{\rm max})/2$ 
	\STATE $r_{1(i+1)} \gets r_{1i}$\\
	\STATE $T \gets (b_1 - r_{1i} \xi_i)/r_{1(i+1)}$\\
	\STATE $r_{2i} \gets h_2(E_i/\xi_i,r_{1i})$, $r_{2(i+1)} \gets h_2(E_{i+1}/T,r_{1(i+1)})$\\
	\STATE $\tilde{b}_2=r_{2i} \xi_i + r_{2(i+1)} T$
	\IF{$(b_2 - \tilde{b}_2) > \epsilon$}
		\STATE $r_{1i}^{\rm max} \gets r_{1i}$ \COMMENT{Less than $b_2$ bits are transmitted to the $2^{nd}$ user. In order to make those bits closer to $b_2$, we need to decrease energy allocated to the $1^{st}$ user by increasing $T$ while transmitting exactly $b_1$ bits. Reducing $r_{1i}^{\rm max}$ guarantees this operation in the next iterate.}
	\ELSIF{$(\tilde{b}_2 - b_2) > \epsilon$} 
		\STATE $r_{1i}^{\rm min} \gets r_{1i}$ \COMMENT{More than $b_2$ bits are transmitted to the $2^{nd}$ user. In order to make those bits closer to $b_2$, we need to increase energy allocated to the $1^{st}$ user by decreasing $T$ while transmitting exactly $b_1$ bits. Raising $r_{1i}^{\rm min}$ guarantees this operation in the next iterate.}
	\ELSE
		\RETURN
	\ENDIF
\ENDLOOP
\STATE \algorithmicprocend
\normalsize
\end{algorithmic}
\end{algorithm}

Next, we first establish that the cost function (that is, the completion time T) strictly decreases after each iteration of FlowRight, until it stops. We then show that the algorithm always stops.

\begin{theorem}
\label{thm:flowright_stops}
\begin{emph}
The following statements hold:
\begin{enumerate}
\item \emph{As FlowRight runs, objective function $T$ of Prob. \ref{pr:MultiuserScheduling} strictly decreases} after each iteration. Conversely, if $T$ did not change after an iteration, then FlowRight stopped at the previous one.
\item \emph{FlowRight stops, and returns a sequence $\{r_{1i}^{\infty},r_{2i}^{\infty}\}$.}
\end{enumerate}
\end{emph}
\end{theorem}

{\emph{Proof.}} 
\begin{enumerate}
\item Suppose that we are on the $k^{th}$ iteration of the algorithm, and so far local optimizations have been performed on all epoch pairs up to $(i-1,i)$. During the $(k-1)^{th}$ iteration, local optimization on epochs $(i+1,i+2)$ has updated $(r_{1(i+1)}^{k-1},r_{2(i+1)}^{k-1})$ with the corresponding bits $(b_{1(i+1)}^{(k-1)},b_{2(i+1)}^{(k-1)})$. As the $k^{th}$ iteration progress, local optimization on the epochs $(i-1,i)$ determines $(r_{1(i-1)}^k, r_{2(i-1)}^k)$ and resets $(r_{1i}^{k-1}, r_{2i}^{k-1})$. Suppose that this local optimization results in $T_{\rm min}$ before the end of the $i^{th}$ epoch. Hence in the rest of that epoch, rate pair changes to $(0,0)$, \ie, a gap occurs. Let $b_j^k\triangleq b_{ji}^{(k-1)}+b_{j(i+1)}^{(k-1)},~ j \in \{1,2\}$ be the total bits to be transmitted in epochs $(i,i+1)$. From Lemma~\ref{lmm:ConstantRate}, by using a constant rate pair within the $i^{th}$ epoch, \ie, filling the gap within the epoch, at least the same number of bits can be transmitted to each user as in the original, slotted allocation. As the $k^{th}$ iteration progress, local optimization on the epochs $(i,i+1)$ determines $(r_{1i}^{k},r_{2i}^{k})$ and resets $(r_{1(i+1)}^{k-1},r_{2(i+1)}^{k-1})$ by at least trying to assign a constant rate pair in the $i^{th}$ epoch using the same amount of energy, hence reducing the number of bits to be transmitted in the $(i+1)^{th}$ epoch. Then, transmission surely ends before the end of the $(i+1)^{th}$ epoch, \ie, the gap moves from the $i^{th}$ epoch to the $(i+1)^{th}$. Therefore, after local optimization, the total time to transmit $b_1^k$ and $b_2^k$ bits within the epoch pair $(i,i+1)$ reduces. Continuing this way, the gap propagates to the end of the $k^{th}$ iteration, hence the initial $B_1$ and $B_2$ bits are transmitted by the time $T^k<T^{k-1}$.

To prove the converse claim, suppose that after $k^{th}$ iteration, $T$ did not change. Then, no gap has occurred during local optimizations; otherwise it would have propagated to the last epoch used and hence, reduced $T$. Therefore, performing further iterations can not create any gaps meaning that algorithm has indeed stopped at $(k-1)^{st}$ iteration.

\item FlowRight initially starts from a feasible $T \geq T^{\rm opt}$, which is obviously lower bounded by $T^{\rm opt}$, the unique smallest completion time. From Part-1, $T^k=T(\{r^k_{1i},r^k_{2i}\})$ is a strictly decreasing real sequence bounded below by $T^{\rm opt}\in \Re$, hence the iterations eventually stop. Therefore, the rate pairs $\{r^k_{1i},r^k_{2i}\}$ converge to some final value $\{r_{1i}^{\infty},r_{2i}^{\infty}\}$.\enp
\end{enumerate}

Now, we show in the following theorem that the schedule returned by FlowRight possesses the properties of OPT listed in Theorem~\ref{thm:OptPowerAllocation} and~\ref{thm:OptRateAllocation}.

\begin{theorem}
\label{thm:flowright_properties}
\begin{emph}
When FlowRight stops,
\begin{enumerate}
\item \emph{Powers are monotonically nondecreasing, \ie, $P_1 \leq P_2 \leq...\leq P_n$.}
\item \emph{Energy consumed during any constant power allocation band equals the total energy harvested in that band.}
\item \emph{The stronger user's rate is monotone nondecreasing, \ie, $r_{11} \leq r_{12} \leq...\leq r_{1n}$,}
\item \emph{If the stronger user's rate changes at the $(i+1)^{th}$ epoch, the weaker user's rate is zero at the $i^{th}$ epoch,}
\item \emph{The weaker user's rate is monotone nondecreasing, \ie, $r_{21} \leq r_{22} \leq...\leq r_{2n}$.}
\end{enumerate}
\end{emph}
\end{theorem}

{\emph{Proof.}} 
\begin{enumerate}
\item Suppose that we can find two epochs $i$ and $i+1$ with powers such that $P_i>P_{i+1}$. From Lemma~\ref{lmm:ConstantRate}, by bringing power levels closer to each other (this never violates energy causality), further local improvement on these epochs is ensured which contradicts the fact that FlowRight has stopped.
\item Suppose that $P_i=P_s\neq P_{s+1},~s-m \leq i \leq s < n_\infty$ for some band of length $m<s$ such that $\sum_{i=s-m}^{s} E_i^\infty < \sum_{i=s-m}^{s} E_i^0$. Then, from Part-1, we have $P_s < P_{s+1}$. But we can transfer up to $\sum_{i=s-m}^{s} E_i^0-\sum_{i=s-m}^{s} E_i^\infty$ units of energy from epoch $s+1$ to $s$, only improving the schedule (cf. Lemma~\ref{lmm:CloserPowers}). This contradicts the assumption that FlowRight has stopped.
\item Suppose $\exists~i \in \{1,..,n_\infty-1\}~s.t.~r_{1i}>r_{1(i+1)}$. From Part-1 and Lemma~\ref{lmm:CloserRate}, by assigning the average rate $\bar{r_1}=(r_{1i}\xi_i+r_{1(i+1)}\xi_{i+1})/(\xi_i+\xi_{i+1})$ to the \emph{stronger} user in epochs, more bits can be transmitted to the \emph{weaker} user which means further local improvement on these epochs is ensured. This contradicts the assumption that FlowRight stopped.
\item Suppose that the \emph{stronger} user's rate changes at the $(i+1)^{th}$ epoch and $r_{2i}\neq 0$. The \emph{stronger} user's rate can only increase, \ie, $r_{1i}<r_{1(i+1)}$, by Part-3. From Lemma~\ref{lmm:CloserRate}, by bringing $r_{1i}$ and $r_{1(i+1)}$ closer to each other using the energy available for the \emph{weaker} user at the $i^{th}$ epoch, overall transmission duration is decreased which contradicts the assumption that FlowRight stopped.
\item Suppose that $r_{2i}>r_{2(i+1)}$. From Part-5, $r_{1i} = r_{1(i+1)}$ since $r_{2i} \neq 0$. From Part-3, the power is monotone increasing. Since $h_2(P,r)$ is monotone increasing in $P$ by the properties of rate region, $r_{2(i+1)} = h_2(P_{i+1},r_{1i}) \geq h_2(P_i,r_{1i}) = r_{2i}$, which contradicts $r_{2i}>r_{2(i+1)}$.\enp
\end{enumerate}

\iffalse
Before moving on with the last theorem proving that \textbf{FlowRight} solves Problem~\ref{pr:MultiuserScheduling}, we explain the details of Algorithm~\ref{alg:LocalOpt_Causality} (local optimization when energy causality is met) in the following. From Theorem~\ref{thm:flowright_properties}, equalizing the rates of \emph{stronger} user is optimal when it is feasible. Then, given $E_i$, $E_{i+1}$ and $b_j^k$, $j \in \{1,2\}$, $T_{\rm min}$ can be determined iteratively (see  Algorithm~\ref{alg:LocalOpt_Causality}) using Bisection Method. Suppose that the algorithm selects $r_{1i}=r_{1(i+1)}$ such that $b_1^k$ bits are transmitted to the \emph{stronger} user as in the optimal solution. Hence, transmission duration in the $(i+1)^{th}$ epoch is $T=(b_1-r_{1i}\xi_i)/r_{1i}$. Then, objective is to find the single root of the monotone decreasing function $f(r_{1i})=\tilde{b}_2-b_2^k$, where $\tilde{b}_2=h_2(E_i/\xi_i, r_{1i})\xi_i+h_2(E_{i+1}/T, r_{1i}).T$ is the number of bits transmitted to the \emph{weaker} user by the end of transmission. Starting with initial domain interval [0,$r_{1i}^{up}=h_1(E_i/\xi_i,0)$], at each iteration interval is bisected and the subinterval in which the root $T_{\rm min}$ lies is selected. Algorithm converges in the limit to the unique solution as the domain is continuous and can be terminated within a certain arbitrarily small tolerance $\epsilon>0$ in practical implementation.
\fi

\begin{theorem}
\label{thm:flowright_optimal}
\emph{The schedule returned by FlowRight is optimal, \ie,}
$T(\{r_{1i}^{\infty},r_{2i}^{\infty}\})=T^{\rm {opt}}$.
\end{theorem}

{\emph{Proof.}} 
Suppose that FlowRight stops and returns a schedule $\{r_{1i}^{\infty},r_{2i}^{\infty}\} \triangleq S^{\rm fr}$, with completion time $T(\{r_{1i}^{\infty},r_{2i}^{\infty}\})\triangleq T^{\rm fr}$.

\begin{figure}[htpb]
\begin{center}
\includegraphics[scale=0.5]{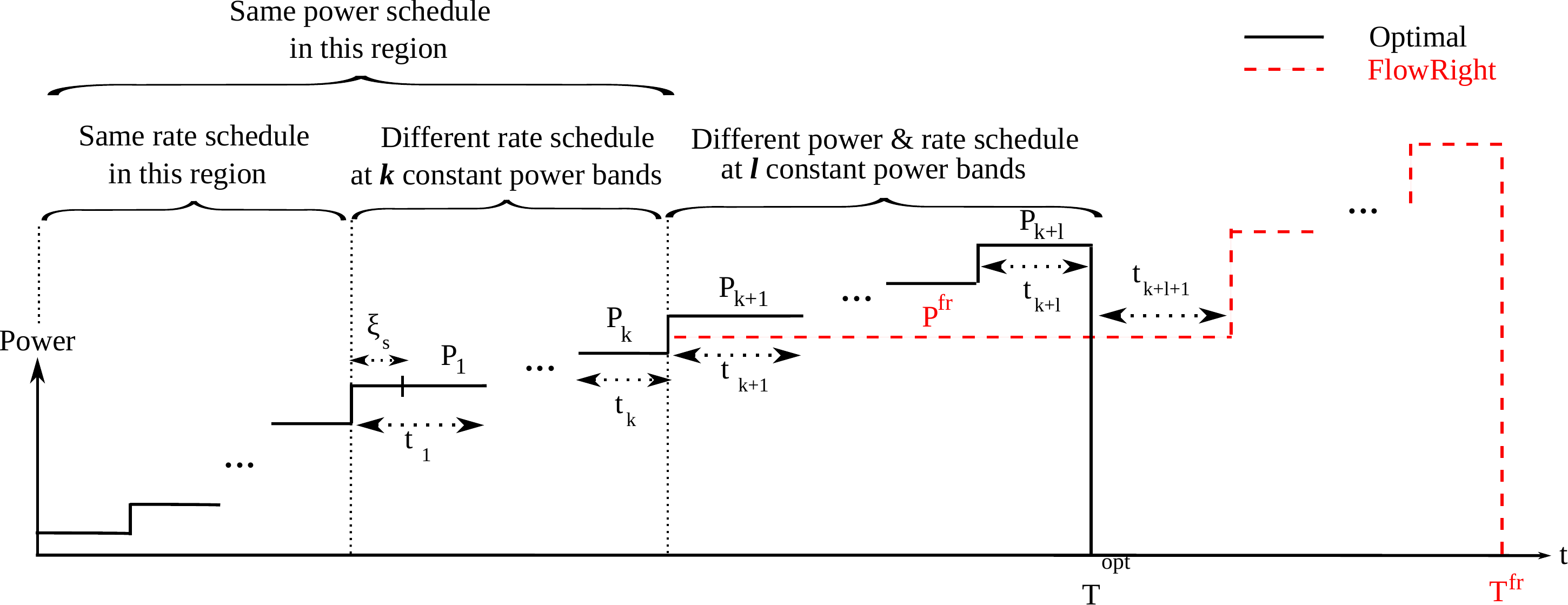} 
\caption{Illustration of the general case (in the proof of Theorem~\ref{thm:flowright_optimal}) that $S^{\rm fr}$ and $S^{\rm opt}$ differ in power at \emph{k} constant power bands and differs in both power and rate at \emph{l} constant power bands.}
\label{fig:flowright_vs_optimal}
\end{center} 
\end{figure}

As FlowRight respects feasibility,  $T^{\rm fr}$ can not be smaller than $T^{\rm opt}$. Suppose $T^{\rm fr} > T^{\rm opt}$. Consider the case that $T^{\rm opt}$ is in the $m^{th}$ epoch and $T^{\rm fr}$ is in the $n^{th}$ epoch with $n \geq m$. % as shown in Fig.~\ref{fig:Theorem_part4}.  
\iffalse
\begin{figure}[htpb]
\begin{center}
\includegraphics[scale=0.6]{Theorem_Part4_Proof.pdf} 
\caption{Illustration of the case $T^{\rm fr} > T^{\rm opt}$ in the proof of Theorem~\ref{thm:flowright_optimal}.}
\label{fig:Theorem_part4}
\end{center}
\end{figure}
\fi
There must be a schedule $\{r_{1i}^{\rm opt},r_{2i}^{\rm opt}\} \triangleq S^{\rm opt}$ that achieves $T^{\rm opt}$. Suppose that $S^{\rm opt}$ and $S^{\rm fr}$ are equal up to epoch $s$, which is the \textit{first} time they differ either in terms of power level or rates, or both. Let us denote the power allocated in epoch $s$ in $S^{\rm opt}$ as $P_{s}^{\rm opt}$ and in $S^{\rm fr}$ as $P_{s}^{\rm fr}$. Consider the following.
\begin{enumerate}
\item $P_{s}^{\rm fr} > P_{s}^{\rm opt}$: From Part-2 of Theorem~\ref{thm:OptPowerAllocation}, all the harvested energies are consumed  within any constant power band of $S^{\rm opt}$. Then, starting from epoch $s$ when $S^{\rm opt}$ consumes all the energy at the end of that constant power region, $S^{\rm fr}$ would have consumed more energy than $S^{\rm opt}$ by Part-1 of Theorem~\ref{thm:flowright_properties}, contradicting the fact that FlowRight always respects energy causality.

\item $P_{s}^{\rm fr} < P_{s}^{\rm opt}$: Suppose that $P_{s}^{\rm fr}$ increases to $P_{s+m}^{\rm fr}$ at some further epoch $s+m$ before $T^{\rm opt}$. From Lemma~\ref{lmm:CloserPowers}, by bringing power levels $P_{s+m-1}^{\rm fr}=P_s^{\rm fr}$ and $P_{s+m}^{\rm fr}$ of the epoch pair $(s+m-1,s+m)$ closer to each other (this never violates energy causality), further local improvement on these epochs is ensured which contradicts the fact that FlowRight has stopped.

\iffalse
\begin{figure}[htpb]
\begin{center}
\includegraphics[scale=0.55]{Theorem_Part4_Proof_2.pdf} 
\caption{Illustration of the case (in the proof of Theorem~\ref{thm:flowright_optimal}) that at the first change between OPT and the schedule returned by FlowRight, power level of OPT is greater than that of the schedule returned by FlowRight.}
\label{fig:Theorem_part4_2}
\end{center} 
\end{figure}
\fi

\end{enumerate}
Hence, $S^{\rm fr}$ cannot have higher power level than $S^{\rm opt}$ until $T^{\rm opt}$. Moreover, if power level of $S^{\rm fr}$ becomes lower than that of $S^{\rm opt}$, then it should stay constant until $T^{\rm opt}$. These results are shown in the general case in Figure~\ref{fig:flowright_vs_optimal}.
\vspace*{-10pt}
Now, suppose that the \emph{first} change occurs when $P_{s}^{\rm fr} = P_{s}^{\rm opt}$ and the rate pairs, $\{r_{1s}^{\rm opt},r_{2s}^{\rm opt}\}$ and $\{r_{1s}^{\rm fr},r_{2s}^{\rm fr}\}$, differ from each other in the general case. Consider the following.
\begin{enumerate}
\item $r_{1s}^{\rm opt} < r_{1s}^{\rm fr}$:  Since $r_{2s}^{\rm opt} = h_2(P_{s}^{\rm opt},r_{1s}^{\rm opt}) > h_2(P_{s}^{\rm fr},r_{1s}^{\rm fr}) \geq 0$, rate of \emph{stronger} user in the $S^{\rm opt}$ should stay constant after epoch $s$ by Theorem~\ref{thm:OptRateAllocation}. Since $T^{\rm fr} \geq T^{\rm opt}$ and $r_{1s}^{\rm fr}\leq r_{1u}^{\rm fr}~\forall u \in~\{s+1,...,n\}$, $S^{\rm fr}$ should have transmitted more bits to \emph{stronger} user than $S^{\rm opt}$, which contradicts the fact that FlowRight always respects \emph{bit feasibility}, \ie, $S^{\rm fr}$ transmits exactly the same number of bits to each user as $S^{\rm opt}$ by the time $T^{\rm fr}$.
\item $r_{1s}^{\rm opt} > r_{1s}^{\rm fr}$: We have $r_{2s}^{\rm fr} = h_2(P_{s}^{\rm fr},r_{1s}^{\rm fr}) > h_2(P_{s}^{\rm opt},r_{1s}^{\rm opt}) \geq 0$; therefore, $r_{1u}^{\rm fr}=r_{1s}^{\rm fr}~\forall u \in~\{s+1,...,m\}$ by Theorem~\ref{thm:flowright_properties}. Then, bit feasibility requires $r_{1}^{\rm fr}\left(\sum_{i=1}^{k+l+1}t_i\right) \leq \sum_{i=1}^{k+l}t_i r_{1(i)}^{\rm opt}$, where $r_{1(i)}^{\rm opt} \geq r_{1s}^{\rm opt}$ is the rate of \emph{stronger} user for the $i^{th}$ constant power band whereas $t_i$ is the duration of that band. Rearranging the terms we have $r_{1}^{\rm fr}\leq \sum_{i=1}^{k+l}\gamma_i r_{1(i)}^{\rm opt}$, where $\gamma_i= t_i \big/ \sum_{i=1}^{k+l+1}t_i \in (0,1) $, $\forall i \in \{1,2,..,k+l\}$. Moreover, from Part-2 of Theorem~\ref{thm:flowright_properties} we have $P^{\rm fr} \left( \sum_{i=k+1}^{k+l+1} t_{i} \right) \geq \sum_{i=k+1}^{k+l} t_{i}P_{i}$, where $P_i$ is the power of the $i^{th}$ constant power band (see Fig.\ref{fig:flowright_vs_optimal}). Rearranging the terms we have $P^{\rm fr} \geq \beta \sum_{i=k+1}^{k+l} (\alpha_{i}/\beta)P_{i}$ where $\alpha_{i}= t_i \big/ \sum_{i=k+1}^{k+l+1}t_{i} \in (0,1)$, $\forall i \in \{1,2,..,k+l\}$ and $\beta = \sum_{i=k+1}^{k+l} t_{i} \big/ \sum_{i=k+1}^{k+l+1}t_i \in (0,1)$. Now, let $\tilde{b}_2^{\rm fr}$ and $\tilde{b}_2^{\rm opt}$ be the number of bits transmitted to the $2^{nd}$ user from epoch $s$ till $T^{\rm opt} + t_{k+l+1}$ by $S^{\rm fr}$ and till $T^{\rm opt}$ by $S^{\rm opt}$, respectively. Then, we have 
\small
\begin{eqnarray}
\tilde{b}_2^{\rm fr}-\tilde{b}_2^{\rm opt} &=&\sum_{i=1}^{k} t_i h_2(P_i, r_1^{\rm fr}) + \left(\sum_{i=k+1}^{k+l+1}t_i\right) h_2(P^{\rm fr}, r_1^{\rm fr}) - \sum_{i=1}^{k+l} t_i h_2(P_i, r_{1(i)}^{\rm opt})\nonumber\\
&=&\left(\sum_{i=k+1}^{k+l+1}t_i\right) \left(\sum_{i=1}^{k} \alpha_i h_2(P_i, r_1^{\rm fr}) + h_2(P^{\rm fr}, r_1^{\rm fr}) - \sum_{i=1}^{k+l} \alpha_i h_2(P_i, r_{1(i)}^{\rm opt})  \right)\nonumber\\
&=&\left(\sum_{i=k+1}^{k+l+1}t_i\right) \left(\sum_{i=1}^{k} \alpha_i \left(\underbrace{h_2(P_i, r_1^{\rm fr})-h_2(P_i, r_{1(i)}^{\rm opt}}_{>0}) \right)+h_2(P^{\rm fr}, r_1^{\rm fr})-\sum_{i=k+1}^{k+l}\alpha_i h_2(P_i, r_{1(i)}^{\rm opt})\right)\nonumber\\
&>& \left(\sum_{i=k+1}^{k+l+1}t_i\right) \left(h_2(P^{\rm fr}, r_1^{\rm fr})-\sum_{i=k+1}^{k+l}\alpha_i h_2(P_i, r_{1(i)}^{\rm opt})\right)\label{eqn:BitFeasibility_1}\\
&\geq & \left(\sum_{i=k+1}^{k+l+1}t_i\right) \left(h_2(\beta \sum_{i=k+1}^{k+l} (\alpha_i/\beta)P_i,r_1^{\rm fr})- \sum_{i=k+1}^{k+l} \alpha_i h_2(P_i,r_{1(i)}^{\rm opt})\right)\label{eqn:BitFeasibility_2}\\
&>& \left(\sum_{i=k+1}^{k+l+1}t_i\right) \left(\beta h_2(\sum_{i=k+1}^{k+l} (\alpha_i/\beta)P_i,r_1^{\rm fr})- \sum_{i=k+1}^{k+l} \alpha_i h_2(P_i,r_{1(i)}^{\rm opt})\right)\label{eqn:BitFeasibility_3}\\
&>& \left(\sum_{i=k+1}^{k+l+1}t_i\right) \left(\beta \sum_{i=k+1}^{k+l} (\alpha_i/\beta) h_2(P_i,r_1^{\rm fr})- \sum_{i=k+1}^{k+l} \alpha_i h_2(P_i,r_{1(i)}^{\rm opt})\right)\label{eqn:BitFeasibility_4}\\
&=&\left(\sum_{i=k+1}^{k+l+1}t_i\right) \left( \sum_{i=k+1}^{k+l} \alpha_i \left( \underbrace{h_2(P_i,r_1^{\rm fr})-h_2(P_i,r_{1(i)}^{\rm opt})}_{>0}\right)\right)~~> 0\label{eqn:BitFeasibility_5}
\end{eqnarray}

\normalsize
Eq~\eqref{eqn:BitFeasibility_1}, Eq~\eqref{eqn:BitFeasibility_2} and Eq~\eqref{eqn:BitFeasibility_5} follows from the $2^{nd}$ property of the rate region while Eq~\eqref{eqn:BitFeasibility_3} and Eq~\eqref{eqn:BitFeasibility_4} follows from the $3^{rd}$. Hence, $S^{\rm fr}$ transmits more bits to the $2^{nd}$ user than $S^{\rm opt}$, contradicting the fact that FlowRight always respects bit feasibility.
\end{enumerate}
\iffalse
\begin{figure}[htpb]
\begin{center}
\includegraphics[scale=0.5]{Theorem_Part4_Proof_4.pdf} 
\caption{Illustration of the case (in the proof of Theorem~\ref{thm:flowright_optimal}) that ---}
\label{fig:Theorem_part4_4}
\end{center} 
\end{figure}
\fi

Then, power allocation and rate pairs of $S^{\rm opt}$ and  $S^{\rm fr}$ cannot differ, so $S^{\rm fr}=S^{\rm opt}$ and $T^{\rm fr}=T^{\rm opt}$.\enp

\section{Algorithm Complexity}
\label{sec:AlgorithmComplexity}

The core computational step in the algorithm is the local optimization (Algorithm~\ref{alg:LocalOpt}). This entails the solution of a nonlinear equation, which in our numerical computations has been done iteratively using the bisection method (detailed in Algorithms~\ref{alg:Find_Tmin} and~\ref{alg:LocalOpt_Causality}). The exact number of iterations, hence the convergence rate depends on the selected tolerance level $\epsilon$, but in our experiments convergence time of a local computation is typically on the order of milliseconds. 

Maximum number of iterations of Algorithm~\ref{alg:Find_Tmin} is proportional to $log_2(T_{\rm upper}/\epsilon_1)$ whereas that of Algorithm~\ref{alg:LocalOpt_Causality} is proportional to $log_2(r_{1i}^{\rm up}/\epsilon_2)$.  Then, the worst case computation time of local optimization be $C_{\rm local}~\propto~\max\{log_2(T_{\rm upper}/\epsilon_1),log_2(r_{1i}^{\rm up}/\epsilon_2)\}$. For $n$ epochs, $n-1$ local optimizations are performed at each iteration. Hence, the worst case computation time of an iteration is $C_{iter}\propto C_{\rm local} \times n^{\rm opt}$.

While FlowRight theoretically terminates when $T$ does not change from one iteration to the next, in terms of implementation it would make sense to stop the iterations when the change is within some $\epsilon$ of the average epoch size. The choice of $\epsilon$ will determine the number of iterations and hence the linear scaling coefficient of complexity. In our extensive simulations based on randomized energy harvest amounts, epoch durations and bits to be transmitted, we observed that the number of iterations for achieving convergence sufficient for all practical purposes is on the order of ${(n^{\rm opt})}^2$. Hence, the worst case computation time of FlowRight is $C_{\rm fr} \propto C_{\rm local} \times {(n^{\rm opt})}^3$ which has polynomial time complexity in $n^{\rm opt}$.  

\section{A Numerical Example}
\label{sec:Example}
Consider a two-user  AWGN broadcast channel with bandwidth $100$KHz and noise power spectral density $N_0=10^{-13}$ Watts/Hz. Suppose that the path loss from sender to the \emph{stronger} and \emph{weaker} users are 70dB and 75dB, respectively. The sender needs to transmit $800$Mbits to the \emph{stronger} user and $100$Mbits to the \emph{weaker} user. Energy harvests of amounts [10, 10, 20, 40, 60, 70, 90, 180, 190, 100, 50, 30, 10] Joules arrive at [0, 2, 5, 7, 9, 10, 11, 13, 14, 15, 18, 20, 23] hours.

\begin{figure}[htpb]
\centering \includegraphics[scale=0.7]{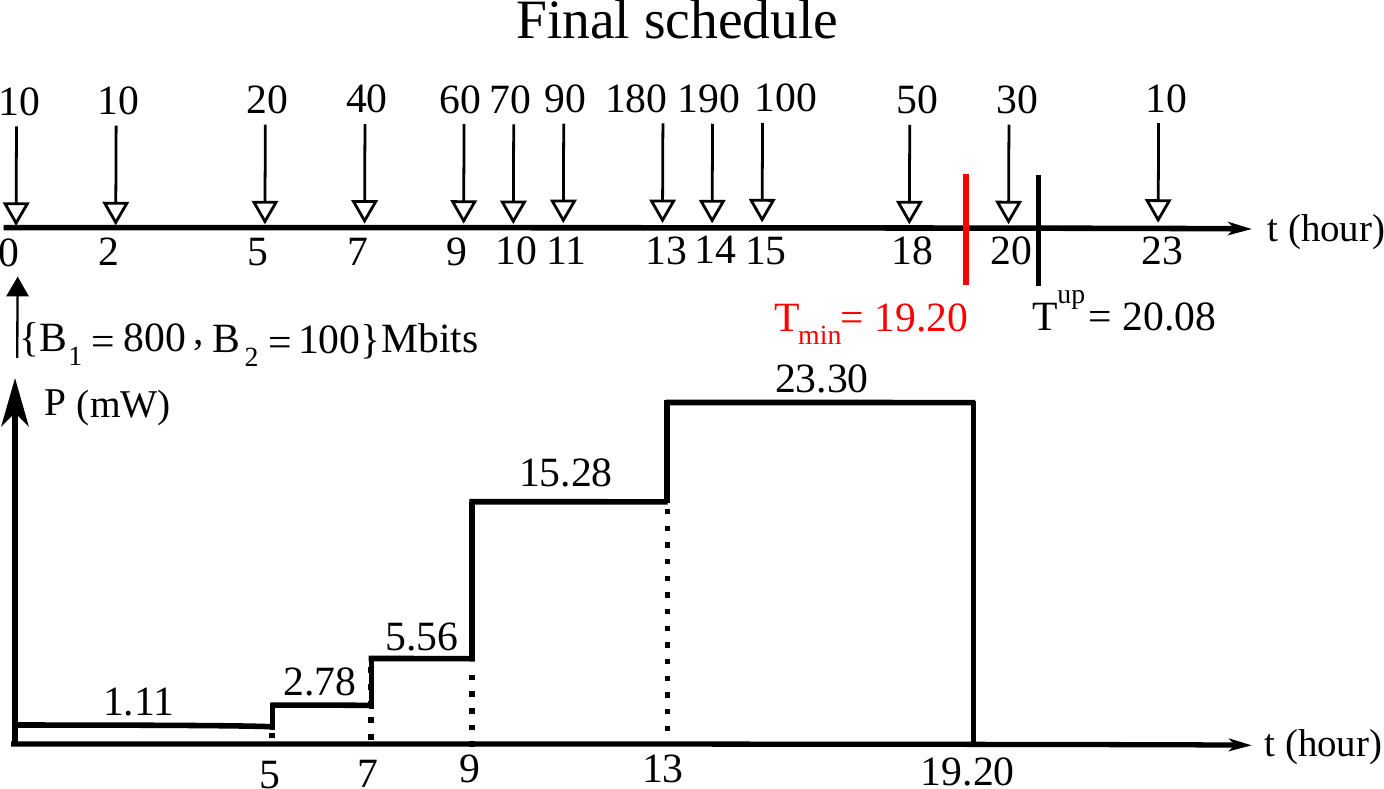}
\caption{A numerical example for the execution of FlowRight. The top figure represents the initial transmission completion time, $T^{\rm up}$=20.08 hours, after the initialization phase and the final transmission completion time, $T_{\rm min}$=19.20 hours after the termination of the algorithm to transmit $B_1$=800Mbits and $B_2$=100Mbits for the given energy harvest instants with the corresponding energy amounts. In the bottom figure, transmit powers are shown to be  [1.11, 2.78, 5.56, 15.28, 23.30] mW for the durations [5, 2, 2, 4, 6.20] hours in the final schedule. The final schedule is $\{(r_{1i}^{\infty},r_{2i}^{\infty})\}$=[(1.6, 0),(1.6, 0),(4.0, 0),(7.8, 0),(18.7, 0.6),(18.7, 0.6),(18.7, 0.6),(18.7, 4.1),(18.7, 4.1),(18.7, 4.1),(18.7, 4.1)]Kbps with durations $\{\xi_i\}$ as shown in the last figure.} 
\label{fig:NumEx}
\end{figure}

FlowRight computes the final schedule shown in Fig. \ref{fig:NumEx}. The stopping criterion used in this example was successive iterations being within $\frac{\epsilon}{n}\sum_{i=1}^{n}\xi_i$ where $\epsilon=10^{-9}$. Running on a PC with Intel Core 2 Duo CPU (2.26GHz) and 2GB RAM, the algorithm stopped after 59 iterations in 1.29 second. In the resulting optimal schedule, the last two energy harvests are not used. Note that in the final schedule, transmit powers remain constant during epochs, and are non-decreasing in time. 
%Also note that bursts of energy harvests that arrive close together are combined and transmit power is kept constant as much as possible. 
 
\section{Conclusions and Future Work}

In this paper, we formulated and solved the offline transmission completion time minimization problem on an energy harvesting broadcast link.  We have observed that, in the optimal solution, energy harvests may not necessarily be depleted at the end of each epoch, and could be deferred for later use. The schedule tries to ``hurry up and be lazy'' at the same time. The sender picks rates from the broadcast capacity region judiciously, such that it completes transmission to both users at the same time, $T$. In the optimal schedule, the powers are non-decreasing in time, so that transmission rate is highest toward the end. We have shown that the problem can be solved efficiently with a modification of the FlowRight algorithm. Our proposed algorithm starts with an upperbound on $T$ and strictly improves it after every iteration or ``pass" through the schedule, and stops when $T$ converges to the optimal value. 

There are a number of directions for further work related to the problem presented in this paper. One of these is solving the offline minimization problem when data arrive during transmission, rather than being available in the beginning. Our preliminary work on this modification of the problem indicates that its solution has similar structural properties to the first problem, such as the powers being nondecreasing in time, and rates not changing between data arrival or energy harvest instants. Here, the optimal solution has more reason to be ``lazy" in terms of transmission rate, as data will continue to come and it may be wise to save energy for future data arrivals. We believe that a further modified version of the iterative algorithm described in this paper solves this version of the problem.

A second direction for further work is addressing the multiple-access version of this problem. There, energy harvests will be occurring at the senders, possibly at different points in time. Finding a distributed solution for that case may be a difficult yet interesting problem. 

Finally, another issue of interest is time-varying channel gain. The case of time-varying channel gain is interesting, and perhaps more meaningful to be setup as an online problem, rather than an offline problem, as channel gain variation is often difficult to predict (whereas energy harvesting times or packet formation times may be known ahead of time in some applications.) As the offline problem formulation has facilitated the analysis of the problem, going to an online formulation is arguably the most important challenge. While there are different ways to formulate the online problem, for example, as a dynamic control problem, our intuition is that approximate methods that leverage the offline formulation may be more tractable and insightful.  
\label{conclusion}

\section{Acknowledgements}
We thank S. Ulukus and K. Leblebicioglu for useful discussions.

\section{Appendix}
\label{subsec:AWGN_h2}

\subsection{Proof of Proposition~\ref{prop:h}}
$1^{st}$ and $2^{nd}$ order partial derivatives of $h_1(P,r)$ and $h_2(P,r)$ for the AWGN BC are as follows:
\small
\begin{eqnarray}
\frac{\partial h_1(P,r)}{\partial P}=\frac{1}{2} (\log_2 e) \frac{s_1s_2}{s_1 s_2 P + s_1 \sigma^2 - (s_1 - s_2) \sigma^2 2^{2r}} \geq 0 && 
\frac{\partial h_2(P,r)}{\partial P}=\frac{1}{2} (\log_{2}e) \frac{s_2}{s_2 P+\sigma^2} \geq 0\label{eq:app1_1}\\
\frac{\partial h_1(P,r)}{\partial r}=-\frac{s_1s_2 P + s_1\sigma^2}{s_1 s_2 P + s_1 \sigma^2 - (s_1 - s_2) \sigma^2 2^{2r}} \leq 0 && 
\frac{\partial h_2(P,r)}{\partial r}=-\frac{2^{2 r}}{(2^{2 r}-1)+\frac{s_1}{s_2}} \leq 0\label{eq:app1_2}\\
\frac{\partial^2 h_1(P,r)}{\partial P^2}=-\frac{1}{2} (\log_2 e) \frac{(s_1s_2)^2}{(s_1 s_2 P + s_1 \sigma^2 - (s_1 - s_2) \sigma^2 2^{2r})^2} \leq 0 &&
\frac{\partial^2 h_2(P,r)}{\partial P^2}=-\frac{1}{2} (\log_{2}e) \frac{{s_2}^2}{(s_2 P+\sigma^2)^2} \leq 0\label{eq:app1_3}\\
\frac{\partial^2 h_1(P,r)}{\partial r^2}=-\frac{(2ln2)(s_1-s_2)(s_1s_2P+s_1\sigma^2)\sigma^2 2^{2r}} {(s_1 s_2 P + s_1 \sigma^2 - (s_1 - s_2) \sigma^2 2^{2r})^2} \leq 0 && \frac{\partial^2 h_2(P,r)}{\partial r^2}=-\frac{(2\ln 2) 2^{2 r} \frac{s_1-s_2}{s_2}}{((2^{2 r}-1)+\frac{s_1}{s_2})^2} \leq 0\label{eq:app1_4}\\
\frac{\partial^2 h_1(P,r)}{\partial r\partial P}=\frac{s_1s_2(s_1-s_2)\sigma^2 2^{2r}}{(s_1s_2P+s_1\sigma^2-(s_1-s_2)\sigma^2 2^{2r})^2}\geq 0 && \frac{\partial^2 h_2(P,r)}{\partial r \partial P}=0\label{eq:app1_5}\\
\frac{\partial^2 h_1(P,r)}{\partial P\partial r}=-\frac{s_1s_2(s_1-s_2)\sigma^22^{2r}}{(s_1s_2P+s_1\sigma^2-(s_1-s_2)\sigma^2 2^{2r})^2} \leq 0
&& \frac{\partial^2 h_2(P,r)}{\partial P \partial r}=0 \label{eq:app1_6}
\end{eqnarray}
\normalsize

From \eqref{eq:Bcast_R1} and \eqref{eq:Bcast_R2}, $h_1(P,r)$ and $h_2(P,r)$ are nonnegative. Monotonicity follows from \eqref{eq:app1_1} and \eqref{eq:app1_2} as the signs of the first order partial derivatives of $h_1(P,r)$ and $h_2(P,r)$ in their respective domains are always fixed. From \eqref{eq:app1_3} and \eqref{eq:app1_4}, $h_1(P,r)$ and $h_2(P,r)$ are concave in power and, respectively rate, when the other parameter is held constant. The last property follows from \eqref{eq:app1_5} and \eqref{eq:app1_6}.

\subsection{Proof of Lemma \ref{lmm:CloserPowers}}
\label{subsec:Appendixbeta} 
Using \eqref{eq:Del_P}, \eqref{eq:f} can be written as
\small
\begin{equation}
f_1(\beta)=h_2(P_1+(1-\beta)\Delta P,\bar{r_{1}})\beta + h_2(P_2-\beta \Delta P,\bar{r_{1}})(1-\beta) - h_2(P_1,r_{11})\beta - h_2(P_2,r_{12})(1-\beta).
\end{equation}
\normalsize
The $2^{nd}$ order derivative of $f_1$ with respect to $\beta$ is  the following~\footnote{$h_{2_{x}}$ and $h_{2_{y}}$ represent the first order partial derivatives of $h_{2}$ with respect to $P$ and $r$, respectively. Second order partial derivatives of $h_{2}$ are represented by $h_{2_{xx}}$, $h_{2_{xy}}$, $h_{2_{yx}}$ and $h_{2_{yy}}$}
\small
\begin{eqnarray}
\frac{\partial ^2 f_1}{\partial^2 \beta} &=&2\{\underbrace{h_{2_{x}}(P_1+(1-\beta)\Delta P,\bar{r_{1}})(-\Delta P)-h_{2_{x}}(P_2-\beta \Delta P,\bar{r_{1}}) (-\Delta P)}_{\leq 0}\}\nonumber\\
&&+2\{\underbrace{h_{2_{y}}(P_1+(1-\beta)\Delta P,\bar{r_{1}})(r_{11}-r_{12})-h_{2_{y}}(P_2-\beta \Delta P,\bar{r_{1}})(r_{11}-r_{12})}_{=0}\}\nonumber
\end{eqnarray}
\begin{eqnarray}
&&+\beta \{ \underbrace{h_{2_{xx}}(P_1+(1-\beta)\Delta P,\bar{r_{1}})(-\Delta P)^2}_{\leq 0} + \underbrace{h_{2_{xy}}(P_1+(1-\beta)\Delta P,\bar{r_{1}})(-\Delta P) (r_{11}-r_{12})}_{=0}\}\nonumber\\
&&+\beta \{ \underbrace{h_{2_{yx}}(P_1+(1-\beta)\Delta P,\bar{r_{1}})(-\Delta P)(r_{11}-r_{12})}_{=0} + \underbrace{h_{2_{yy}}(P_1+(1-\beta)\Delta P,\bar{r_{1}})(r_{11}-r_{12})^2}_{\leq 0}\}\nonumber \\
&&+(1-\beta)\{ \underbrace{h_{2_{xx}}(P_2-\beta \Delta P,\bar{r_{1}})(-\Delta P)^2}_{\leq 0} + \underbrace{h_{2_{xy}}(P_2-\beta \Delta P,\bar{r_{1}})(-\Delta P)(r_{11}-r_{12})}_{=0}\}\nonumber \\
&&+(1-\beta)\{\underbrace{h_{2_{yx}}(P_2-\beta \Delta P,\bar{r_{1}})(-\Delta P)(r_{11}-r_{12})}_{=0} + \underbrace{h_{2_{yy}}(P_2-\beta \Delta P,\bar{r_{1}})(r_{11}-r_{12})^2}_{\leq 0}\} ~~\leq 0\label{eq:fxx} 
\end{eqnarray}
\normalsize

According to the properties of the rate region (1)-(4), \eqref{eq:fxx} always holds. Hence $f_1$ is concave in $\beta$. \enp

\subsection{Proof of Lemma \ref{lmm:CloserRate}}
\label{subsec:Appendix_closer_rates} 
Substituting $r_{11}^{'}= r_{11} + (1-\beta)\Delta r$ and $r_{12}^{'}= r_{12}-\beta \Delta r$ in to \eqref{eq:rate_f}, we have the following.
\small
\begin{eqnarray*}
f_2(\beta)&=& h_2(P_1,r_{11} + (1-\beta)\Delta r)\beta + h_2(P_2,r_{11}-\beta \Delta r)(1-\beta) - h_2(P_1,r_{11})\beta - h_2(P_2,r_{12})(1-\beta).
\end{eqnarray*}
\normalsize
The $2^{nd}$ order derivative of $f_2$ with respect to $\beta$ is the following
\small
\begin{eqnarray}
\frac{\partial^2 f_2}{\partial \beta^2} &=&2\{\underbrace{h_{2_{y}}(P_1,r_{11} + (1-\beta)\Delta r)(-\Delta r) - h_{2_{y}}(P_2,r_{12}-\beta\Delta r)(-\Delta r)}_{\leq 0}\} \nonumber \\
&& + \{\underbrace{\beta(\Delta r)^2 h_{2_{yy}}(P_1,r_{11} + (1-\beta)\Delta r)}_{\leq 0} + \underbrace{(1-\beta)(\Delta r)^2 h_{2_{yy}}(P_2,r_{12} - \beta \Delta r)}_{\leq 0}\} ~~\leq 0\label{eq:rate_fxx}
\end{eqnarray}
\normalsize
According to the properties of the rate region (1)-(4), \eqref{eq:rate_fxx} always holds. Hence $f_2$ is concave in $\beta$. \enp

\singlespacing

\end{document}